\documentclass{article}
\usepackage{arxiv}
\usepackage[utf8]{inputenc} 
\usepackage[T1]{fontenc}    
\usepackage{hyperref}       
\usepackage{url}            
\usepackage{booktabs}       
\usepackage{amsfonts}       
\usepackage{nicefrac}       
\usepackage{microtype}      
\usepackage{lipsum}		
\usepackage{xcolor}
\usepackage{graphicx}
\usepackage{subcaption}
\usepackage{wrapfig}
\usepackage{rotating}
\usepackage{adjustbox}
\usepackage{enumitem}
\usepackage[autostyle=true]{csquotes} \usepackage[xindy]{glossaries}
\DeclareUnicodeCharacter{FB01}{fi}
\makeglossaries
\setglossarystyle{listgroup}

\newglossaryentry{TLP}
{
  name={TLP},
  description={The Traffic Light Protocol (TLP) was created to facilitate information sharing. TLP is a set of designations used to ensure that sensitive information is shared with the appropriate audience. It uses four colours to indicate the expected sharing limits to be applied by the beneficiary(ies). TLP has only four colours; all designations not listed in this standard are not considered valid by FIRST. Translated from \cite{TrafficLightProtocol}}
}
\newglossaryentry{Sand-box}
{
  name={Sand-box},
  description={A Sand-Box is a security mechanism to separate programs running, usually to mitigate system failures or software vulnerabilities. It is often used to run untested or unreliable programs or code, possibly from third parties, suppliers, users or websites that are not verified or at risk, without risking harm to the host machine or operating system. Translated from \cite{SandboxComputerSecurity2019}}
}
\newglossaryentry{PFE}
{
  name={PFE},
  description={Research or research and development work, carried out individually or in very small groups over 4 to 6 months of full-time equivalent. The IEPTs take place in a laboratory of INSA Lyon or in a company. In all cases, they are supervised and supervised by INSA Lyon teacher-researchers and give rise to a report and defence. From \cite{ProjetsFinEtude2009}}
}
\newglossaryentry{TLSH}
{
  name={TLSH},
  description={TLSH is a fuzzy-hashing library. From a bytes stream with a minimum length of 50 bytes, TLSH generates a hash value that can be used for similarity comparisons. Similar objects will have similar hash values, which allows similar objects to be detected by comparing their hash values. Translated from \cite{TrendmicroTlsha}}
}
\newglossaryentry{SSDeep}
{
  name={SSDeep},
  description={SSDeep is a hash calculation program of the type "context triggered piecewise hashes" (CTPH). Also called fuzzy hashing, a CTPH algorithm can identify similar entries. These entries are sequences of identical bytes in the same order, although the bytes between these sequences may be different in content and length. Translated from \cite{SsdeepFuzzyHashing}}
}
\newglossaryentry{SIFT}
{
  name={SIFT},
  description={Method of extracting distinctive invariant elements from images that can be used to make a reliable comparison between different views of an object or scene. The characteristics are invariant to the scale and rotation of the image, and have been shown to provide a robust match over a range of sharp distortion, 3D modification, noise addition and lighting modification. Translated from \cite{ScaleinvariantFeatureTransform}}
}
\newglossaryentry{DigInPix}
{
  name={DigInPix},
  description={Automatic system capable of identifying visual entities appearing in images and videos, from a list of 25,000 entities, aggregated from Wikipedia, and more specific websites. DigInPix is a generic application designed to identify different types of entities, designed by and for the INA. Translated from the \cite{DigInInPixVisualNamedEntities2015}}
}
\newglossaryentry{SYVICOL}
{
  name={SYVICOL},
  description={Syndicat des Villes et Communes du Luxembourg \cite{AccueilSyvicol}}
}
\newglossaryentry{SIGI}
{
  name={SIGI},
  description={Intercommunal Computer Management Union \cite{Intercommunal Management Union}}
}

\newglossaryentry{Image-Retrieval}
{
  name={Image-Retrieval},
  description={An image search system is one of the solutions that can be used to navigate, search and retrieve images from a large database of digital images. Translated from \cite{marshallSurveyImageRetrieval}}
}
\newglossaryentry{Image-Matching}
{
  name={Image-Matching},
  description={Image matching is a classic problem in scene analysis: if you give a reference image of an object, you have to decide if it exists in a scene image being analyzed, and find its location if it does. The measurement of similarity between two images is an objective of Image-Matching. Translated from \cite{douRobustImageMatching2018}}
}
\newglossaryentry{Image-Classification}
{
  name={Image-Classification},
  description={Image classification is a process of grouping pixels into several classes based on the application of logical decision rules. Translated from \cite{ImageClassificationOverview}}
}
\newglossaryentry{Open-Set-Classification}
{
  name={Open-Set-Classification},
  description={For some classification problems, we do not know all the possible classes during the labeling phase. An Open-Set-Classificaiton situation is a situation where many labels are known, but more and more other labels are not yet known. Translated from \cite{scheirerOpenSetRecognition2013a}}
}
\newglossaryentry{Redis}
{
  name={Redis},
  description={Redis is an open-source in-memory database (under BSD license), used as a database, cache and message broker. It supports data structures such as strings, hashes, lists, sets, sets, sets sorted with queries by range, bitmaps, geospatial indexes with queries by radius. Redis integrates replication, Lua scripting, transactions and different levels of disk persistence, and offers high availability via Redis Sentinel and automatic partitioning with Redis Cluster. Translated from \cite{Redis}}
}
\newglossaryentry{SMILE}
{
  name={SMILE},
  description={Securitymadein.lu is the main online source for cybersecurity in Luxembourg. Its purpose is to provide news, relevant information and a toolbox with cybersecurity solutions useful for private users, organizations and the community. Translated from \cite{SECURITYMADEINLUa}}
}
\newglossaryentry{CIRCL}
{
  name={CIRCL},
  description={Computer Incident Response Center Luxembourg is a government initiative designed to provide a systematic response to IT security threats and incidents. Translated from \cite{CIRCLMissionStatement}}
}
\newglossaryentry{C3}
{
  name={C3},
  description={CyberSecurity Competence Center offers services in collaboration with partners, promotes the long-term development of wealth and skills while increasing Luxembourg's attractiveness and reputation in the field of cybersecurity. Three main objectives: observe, train and test. Translated from \cite{CybersecurityCompetenceCenter}}
}
\newglossaryentry{CASES}
{
  name={CASES},
  description={CyberSecurity Awareness and Security Enhancement Services is an initiative of the Luxembourg government that provides awareness activities, web resources and other tools such as MONARC, focusing on understanding the information security issues facing SMEs. Translated from \cite{CASESLU}}
}
\newglossaryentry{Forensic}
{
  name={Forensic},
  description={Forensics studies are a branch of digital forensics that focuses on the evidence found in computers and digital storage media. The purpose of forensics is to examine digital media in a legal manner in order to identify, preserve, retrieve, analyze and present facts and opinions about digital information. Translated from \cite{ComputerForensicsWikipedia}}
}

\newglossaryentry{RGPD}
{
  name={RGPD},
  description={The General Data Protection Regulation is a regulation of Community law on data protection and privacy for all citizens of the European Union and the European Economic Area. It also deals with the export of personal data outside the EU and the EEA. Translated from \cite{GeneralDataProtection2019}}
}
\newglossaryentry{ISO 27XXX}
{
  name={ISO 27XXX},
  description={The ISO/IEC 27000 family of standards helps organizations secure their information resources. The purpose of using this family of standards is to standardize the requirements of an information security management system (ISMS). ISO 27005 is risk management oriented. Translated from \cite{ISOIEC27001}}
}
\newglossaryentry{MISP}
{
  name={MISP},
  description={MISP, for Malware (or Magic) Information Sharing Platform, is as its name suggests a platform for sharing security events (broader than just malware).
MISP is an open source software solution that enables the collection, storage, distribution and sharing of cybersecurity indicators and threats related to the analysis of cybersecurity incidents and malware. MISP is designed by and for incident analysts, security and ICT professionals or malware reversers to support their daily operations and effectively share structured information. The objective of the MISP is to promote the sharing of structured information within the IT security community. MISP provides functionalities to support the exchange of information but also the consumption of this information by Network Intrusion Detection Systems (NIDS), LIDS but also log analysis tools, SIEMs. Translated from \cite{MISPMISPMISPISP}\cite{wagnerMISPDesignImplementation2016}}
}
\newglossaryentry{AIL}
{
  name={AIL},
  description={AIL, for Analysis Information Leak framework, is a deep-web crawler, which will analyze the loaded content of the pages to detect possible information leaks. AIL is a modular environment for analyzing potential information leaks from unstructured data sources such as Pastebin texts, similar services or unstructured data flows. The LIA framework is flexible and can be extended to support other functionalities to extract or process sensitive information (e.g. data leak prevention). \cite{CIRCLAILAnalysis} for reference, translated from \cite{AILFrameworkAnalysis2019a}\cite{mokaddemAILDesignImplementation2018}}
}
\newglossaryentry{D4}
{
  name={D4},
  description={D4 is an IoT protocol analyzer that can be easily extended to any parser and/or anaylator. It is a network of widely distributed sensors to monitor DDoS and other malicious activities while remaining an open and collaborative project. D4 is therefore a set of software components, which includes everything you need to create your own sensor network or connect to an existing sensor network using simple clients. From \cite{HomeD4Project} and translated from \cite{D4projectD4coreD4}.}
}
\newglossaryentry{Lookyloo}
{
  name={Lookyloo},
  description={Lookyloo is a web site and content analyzer that is dynamically loaded. Lookyloo is therefore a web interface that allows you to crawl a website and display a tree structure of the domains that are called. Translated from \cite{LookylooWebInterface2019}}
}
\newglossaryentry{URLAbuse}
{
  name={URLAbuse},
  description={URLAbuse is a static website verification tool, which can give an indicative opinion on the probability that the evaluated site is a phishing site. URL Abuse is a free and versatile software for URL verification, analysis and blacklist creation. URLAbuse consists of a web interface where requests are submitted asynchronously and a back-end system to process URLs in a modular way. Translated from \cite{URLAbuseVersatile2019}}
}
\newglossaryentry{Monarc}
{
  name={Monarc},
  description={Developed by \glsenablehyper\gls{CASES}, Monarc \cite{g.i.eMONARC} is a risk assessment and management tool, in conjunction with ISO certifications, standards (including GDPR), etc. The software allows a complete risk analysis to be carried out on the company. The advantage of MONARC is the capitalisation of risk analyses already carried out in similar business contexts: the same vulnerabilities regularly appear in many companies, as they face the same threats and generate similar risks. Most companies have servers, printers, a fleet of smartphones, Wi-Fi antennas, etc. so the vulnerabilities and threats are the same. It is therefore sufficient to generalize the risk scenarios for these assets (also called objects) by context and/or activity. Translated from \cite{MONARCMethodOptimised2019}}
}
\newglossaryentry{MISP-Modules}
{
  name={MISP-Modules},
  description={MISP modules are stand-alone modules that can be used for the expansion and provision of other services. The modules are written in Python 3 following a simple API interface. The objective is to facilitate the extension of MISP functionalities without modifying the basic components. The API is available via a simple REST API that is independent of the installation or MISP configuration. Translated from \cite{ModulesExpansionServices2019}}
}
\newglossaryentry{MISP-Taxonomies}
{
  name={MISP-Taxonomies},
  description={MISP Taxonomies is a set of common classification libraries for tagging, classifying and organizing information. Taxonomy allows the same vocabulary to be expressed among a distributed set of users and organizations. These taxonomies can be used in MISP and other information sharing tools and expressed in labels (triplets). A machine tag is composed of a namespace (MUST), a predicate (MUST) and a value (OPTIONAL). Machine tags are often called triple tags because of their format. Translated from \cite{TaxonomiesUsedMISP2019b}}
}

\newglossaryentry{Carl-Hauser}
{
  name={Carl-Hauser},
  description={Open-Source Framework for testing algorithms for correlation, distance and image analysis. Strongly linked to: Douglas-Quaid. Translated from \cite{OpenSourceTesting2019}}
}
\newglossaryentry{Douglas-Quaid}
{
  name={Douglas-Quaid},
  description={Free software for correlation, distance and image analysis. Strongly linked to: Carl-Hauser. Translated from \cite{OpenSourceSoftware2019a}}
}
\newglossaryentry{VisJS-Classificator}
{
  name={VisJS-Classificator},
  description={Classifier for image matching and grouping. Fast and visual. Uses VISjs and Socket.io to allow the classification/matching of images, manually, in real time in multi-user mode. translated from \cite{vincent-circlClassificatorPicturesMatching2019b}}
}
\newglossaryentry{VisJS-Network}
{
  name={VisJS-Network},
  description={Tool for viewing networks composed of nodes and links. The visualization supports different shapes, styles, colors, formats, images, etc. Network visualization works without problems on any modern browser up to a few thousand nodes and edges. To manage a larger number of nodes, the network has clustering support. The network uses HTML canvas for rendering. The library is easy to modify and already used in MISP in a modified and improved version. Translated from \cite{VisJsNetworka}}
}
\newglossaryentry{Fuzzy-Hash}
{
  name={Fuzzy-Hash},
  description={Signature/hash technique that combines a number of traditional hashes whose limits are determined by the context of the entry. These signatures can be used to identify modified versions of known files even if data has been inserted, modified or deleted in the new files. This method has applications in forensic computer science. Translated from \cite{kornblumIdentifyingAlmostIdentical2006a}}
}
\newglossaryentry{ORB}
{
  name={ORB},
  description={Very fast approach based on binary descriptors based on BRIEF, which is rotationally invariant and resistant to noise. ORB is two orders of magnitude faster than SIFT, while also performing well in many situations. Can be applied to object detection or patch-tracking on a smartphone. translated from \cite{rubleeORBEfficientAlternative2011a}}
}
\newglossaryentry{Feature-Point}
{
  name={Feature-Point},
  description={A Feature-Point is defined as an "interesting" part of an image, and the characteristics are used as a starting point for many Computer Vision algorithms. Since the characteristics are used as the starting point and main primitives for the following algorithms, the processing algorithm will often be as good as its feature-point detector. Translated from \cite{FeatureDetectionComputer2019}}
}
\newglossaryentry{Hamming}
{
  name={Hamming},
  description={Hamming distance is used in telecommunication to count the number of altered bits between two messages of a given length. The distance is compatibilized as the number of differences between the two bit strings. From \cite{DistanceHamming2019}}
}

\newglossaryentry{OpenSource}
{
  name={OpenSource},
  description={Open source is a term indicating that a product includes permission to use its source code, design documents or content. It most often refers to the open-source model, in which software or other products are published under open-source license. Translated from \cite{OpenSourceWikipedia}}
}
\newglossaryentry{OpenData}
{
  name={OpenData},
  description={Open Data is the publication of data that can be used and republished without restrictions of copyright, patents or other control mechanisms. The OpenData initiative is supported by the launch of government initiatives. This data can support the research effort in particular areas by providing an essential raw material for researchers. Translated from \cite{OpenData2019}}
}
\newglossaryentry{Zotero}
{
  name={Zotero},
  description={Zotero is a free tool to help collect, organize, cite and share research sources, such as papers, theses, etc. Translated from \cite{ZoteroYourPersonal}}
}
\newglossaryentry{DataTurks}
{
  name={DataTurks},
  description={Data annotation tool for Machine Learning. The tool has been designed to allow teamwork. After sending the data to the instance and adding team members, it is possible to annotate and create a set of learning or evaluation data in a few hours.
  \cite{dataturksMLDataAnnotations2019} and translated from \cite{dataturksMLDataAnnotations2019a} }
}

\title{Carl-Hauser - Open Source Image Matching Algorithms Benchmarking Framework}

\author{
  Vincent~Falconieri\\
  CIRCL\\
  Luxembourg\\
}

\begin{document}

\maketitle

\begin{abstract}
Security analysts need to classify, search and correlate numerous images. Automatic classification tools improve the efficiency of such tasks. Many \textit{Image-Matching} algorithms are presented in the litterature. The present paper introduces and provides a Open-Source benchmarking and evaluation tool for these algorithms. Is this paper, the framework evaluates algorithms on illustrative datasets, which are constituted of phishing and onion websites. Datasets are provided as Open-Data.
\end{abstract}


\section{Introduction}
CERTs - as CIRCL - and security teams collect and process content such as images (at large from photos, screenshots of websites or screenshots of sandboxes). 
Datasets become larger - e.g. on average 10000 screenshots of onion domains websites are scrapped each day in AIL, an analysis tool of information leak - and analysts need to classify, search and correlate through all the images.

Automatic tools can help them in this task. Less research about image matching and image classification seems to have been conducted  exclusively on websites screenshots. \cite{sampatCNNTaskClassification}\cite{aburrousPredictingPhishingWebsites2010a}\cite{chenFightingPhishingDiscriminative2009}

Our long-term target is to build a generic library and services which can at least be easily integrated in \textit{Threat Intelligence tools} such as 
\textbf{AIL}\footnote{Analysis Information Leak framework - \href{https://github.com/CIRCL/AIL-framework}{github.com/CIRCL/AIL-framework}}\cite{mokaddemAILDesignImplementation2018}  and 
\textbf{MISP}\footnote{Malware Information Sharing Platform - \href{https://github.com/MISP/MISP}{github.com/MISP/MISP}}\cite{wagnerMISPDesignImplementation2016}. A quick-lookup mechanism for correlation would be necessary and part of this library. An evaluation framework is provided as \textit{Carl-Hauser}\footnote{\href{https://github.com/CIRCL/carl-hauser}{github.com/CIRCL/carl-hauser}} and the open-source library itself is provided as \textit{Douglas-Quaid}\footnote{\href{https://github.com/CIRCL/douglas-quaid}{github.com/CIRCL/douglas-quaid}}.

MISP is an open source software solution tool developed at CIRCL for collecting, storing, distributing and sharing cyber security indicators and threats about cyber security incidents analysis. \\
AIL is also an open source modular framework developed at CIRCL to analyze potential information leaks from unstructured data sources or streams. It can be used, for example, for data leak prevention.

Image-matching algorithms benchmarks already exist \cite{gaillardLargeScaleReverse2017}\cite{zaunerImplementationBenchmarkingPerceptual2010}\cite{bianImageMatchingApplicationoriented2017} and are highly informative, but none is delivered turnkey. 

\pagebreak

\subsection{Problem Statement}

\begin{wrapfigure}{r}{0.3\textwidth}
\includegraphics[width=0.3\textwidth]{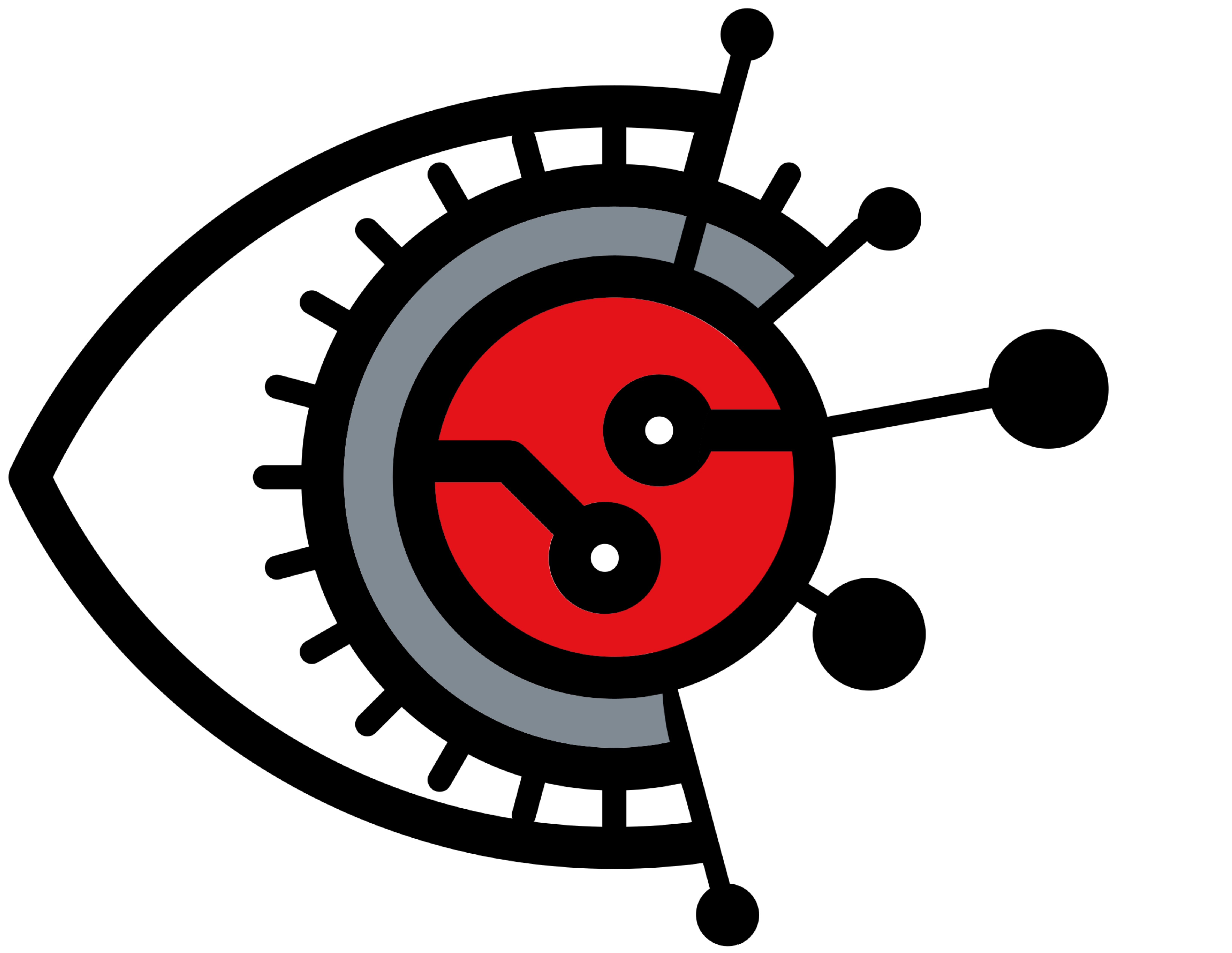} 
\caption{Carl-Hauser logo} 
\end{wrapfigure}

Image correlation for security event correlation purposes is nowadays mainly manual. No open-source tool provides easy correlations, without regard to the technology used. Ideally, the extraction of links or correlation between these images could be fully automated. Even partial automation would reduce the burden of this task on security teams. 

\textbf{The main contribution of this paper is a free and open-source automated bench-marking framework for Image-Matching algorithms review.}

This paper also presents research results for visual clustering of phishing websites.

\section{Datasets}

Tests, performances and speed evaluation were conducted using real sets of pictures. These datasets were extracted from CIRCL's tools, such as \glsenablehyper\gls{AIL} and \glsenablehyper\gls{URLAbuse}. 
One main dataset was used : \href{https://www.circl.lu/opendata/datasets/circl-phishing-dataset-01/}{\textit{circl-phishing-dataset-01}} of 470+ pictures.

Phishing datasets is available for research purposes at \href{https://www.circl.lu/opendata/datasets/circl-phishing-dataset-01/}{\textit{https://www.circl.lu/opendata/datasets/circl-phishing-dataset-01}} 

More details about these datasets can be found at {\color{red}\textbf{ref paper + info link}}

\begin{figure}[h!]
  \centering
  \begin{subfigure}[b]{0.49\linewidth}
    \includegraphics[width=\linewidth]{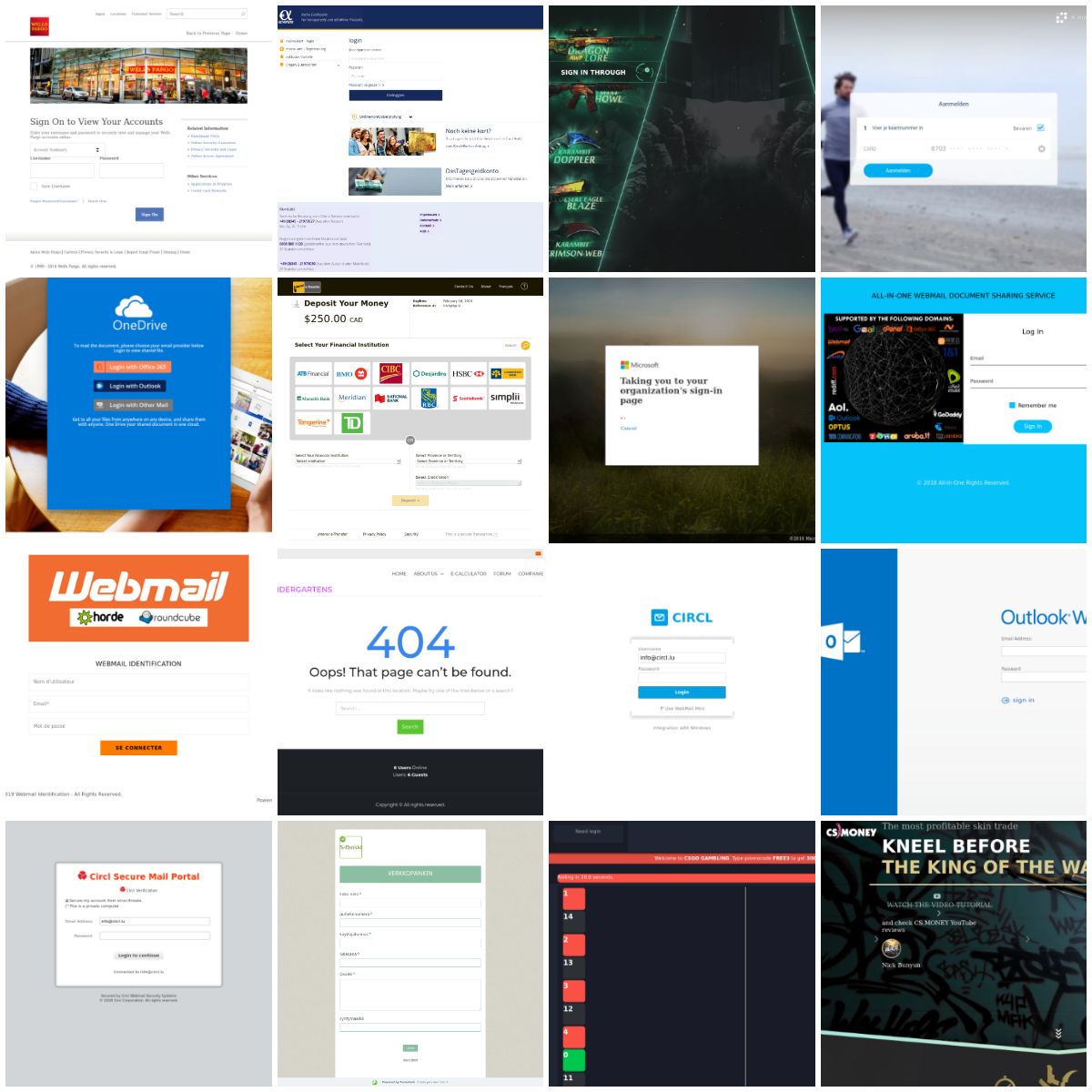}
    \caption{Phishing dataset overview (470+ pictures)}
  \end{subfigure}

  \caption{Dataset's samples}
  \label{fig:datasetsextract}
\end{figure}

\section{Materials and Methods}
In the quest of an use-case-specific Image-matching library, algorithms and approaches are numerous. Benchmarking each one of them is tough and time-consuming. Therefore, we developed a benchmarking framework. The envisioned goal is to allow a \textit{fast implementation} and \textit{fast evaluation} of any new Image-matching library or algorithm, that could come up at a later date. 

\begin{figure}[h!]
  \centering
  \includegraphics[width=\textwidth]{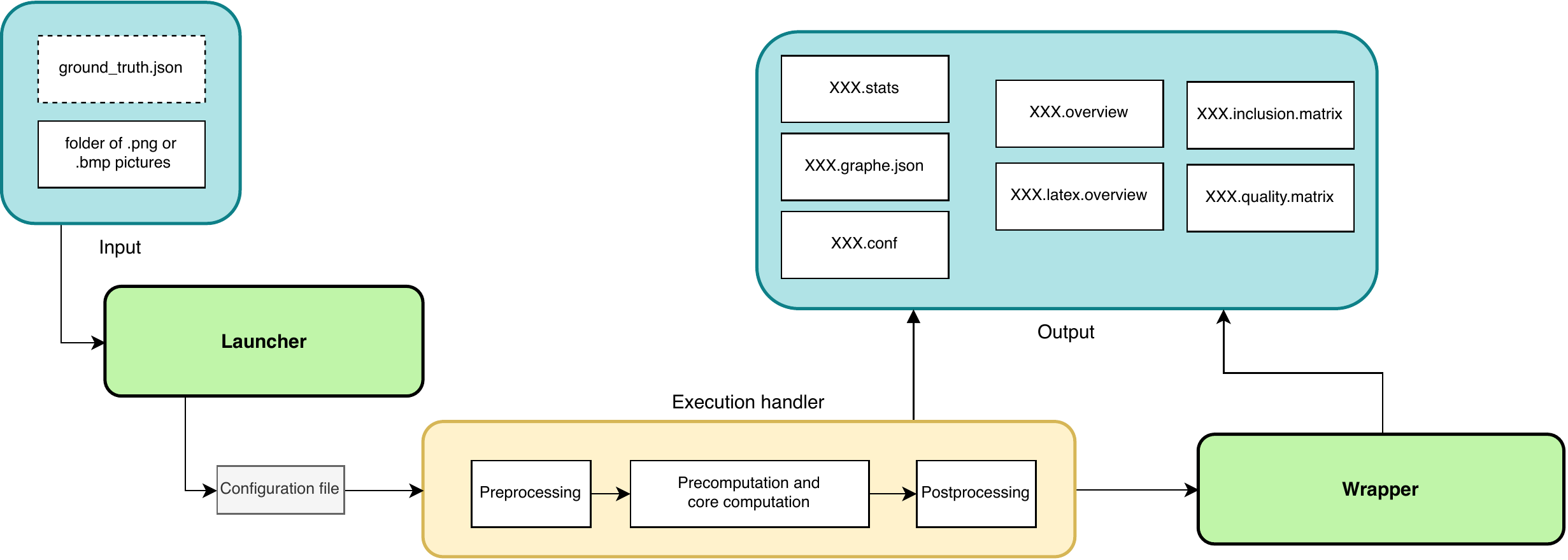}
  \caption{Global overview of benchmarking framework}
  \label{fig:image_matching_pipeline}
\end{figure}

Figure \ref{fig:image_matching_pipeline} presents a global overview scheme of the framework. Given an input folder containing pictures (PNG or BMP), the framework runs all programmed matching algorithms, and returns a series of outputs (quality, timing measure ...). For quality measure, a ground truth file needs to be provided. This file can easily be built with \textbf{VisJS-Classificator}\cite{vincent-circlClassificatorPicturesMatching2019}\footnote{Open-source and built-for-the-occasion manual image classification tool - \href{https://github.com/Vincent-CIRCL/visjs_classificator}{github.com/Vincent-CIRCL/visjs\_classificator}}.

The framework explores the parameters space of the algorithms, by generating configuration files on the go.

These configuration files are read by an execution handler which takes care of the general pipework, including monitoring mechanisms and feeding the configuration file to the core computation handler. This core computation handler is an overwritten version of few abstract methods, specific to used library. 

Pre and post computations can be enabled or disabled and are part of the configuration : OCR, text-hider (Fig.\ref{fig:painted}), image conversion, edge detection ...

\begin{figure}[h!]
  \centering
  \begin{subfigure}[b]{0.4\linewidth}
    \includegraphics[width=\linewidth]{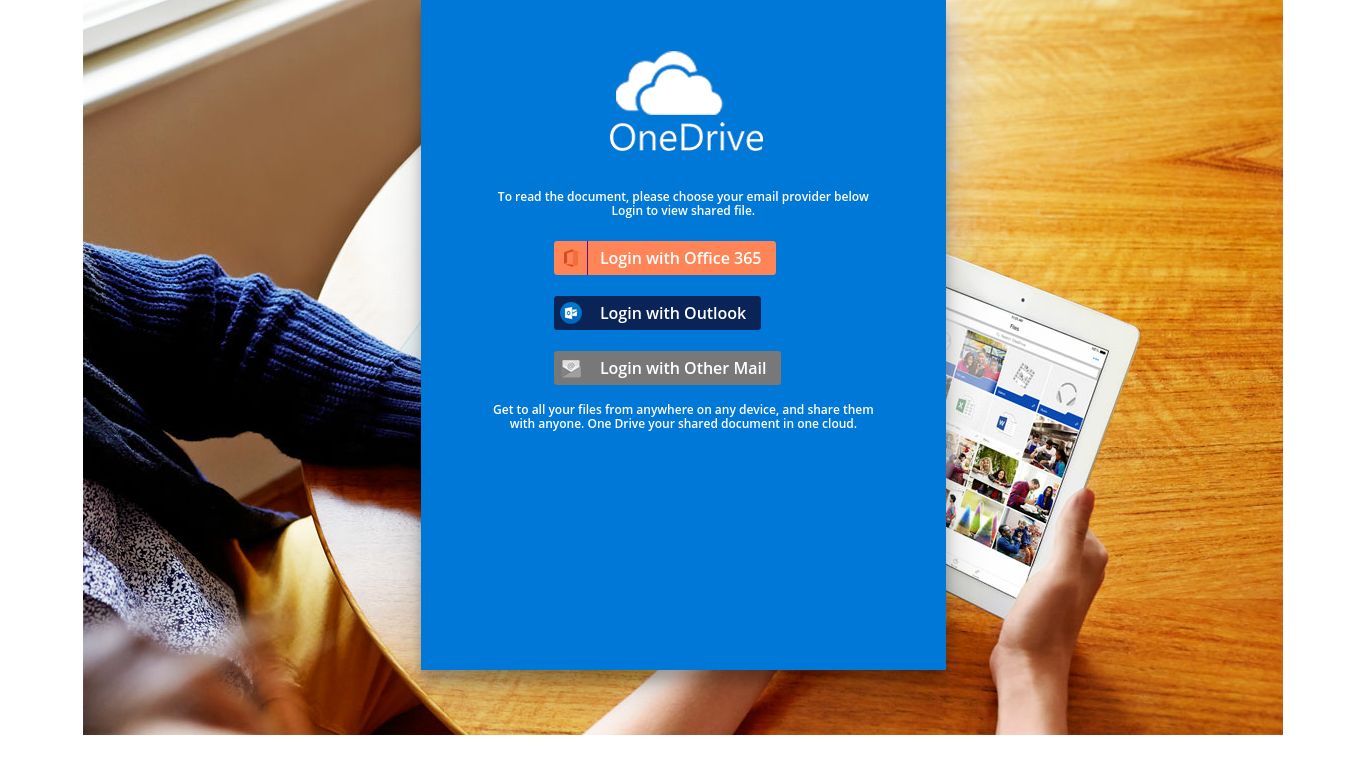}
    \caption{Original picture}
  \end{subfigure}
  \begin{subfigure}[b]{0.4\linewidth}
    \includegraphics[width=\linewidth]{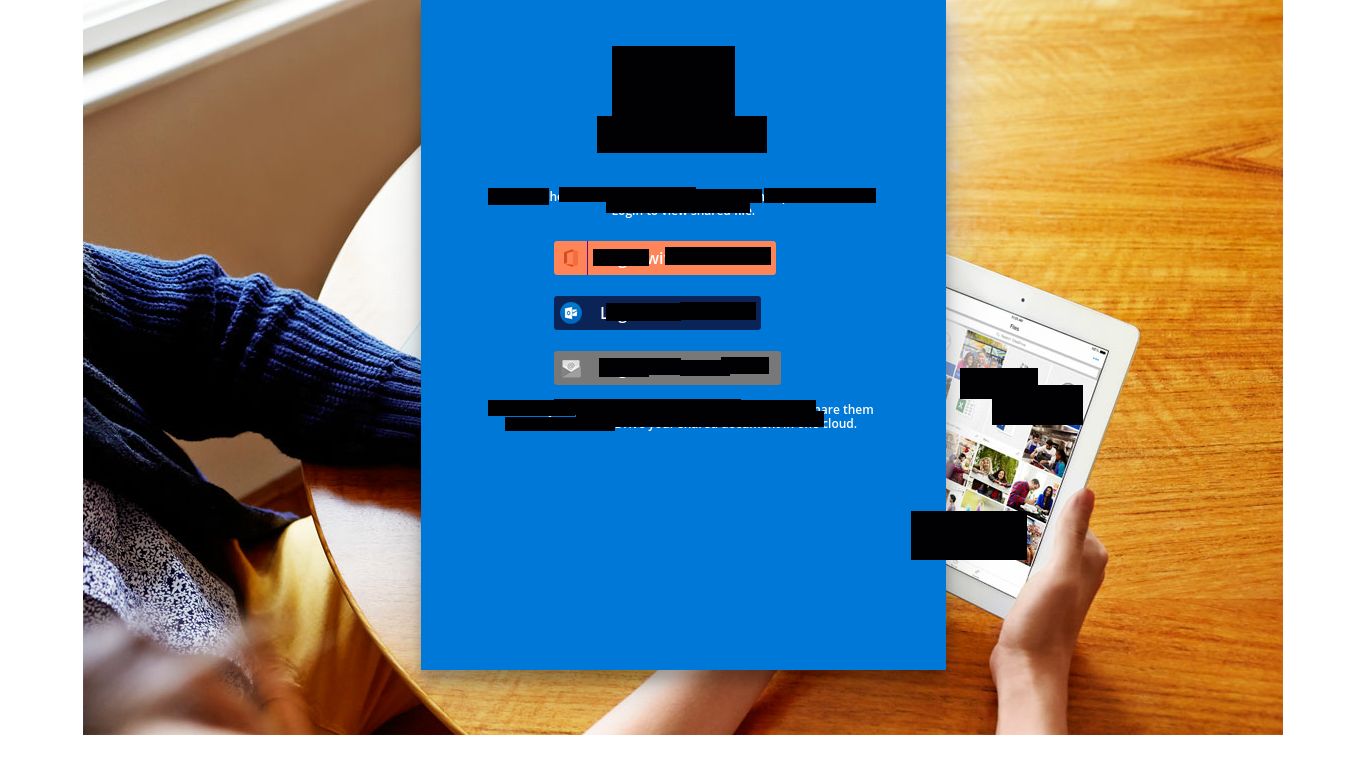}
    \caption{Painted picture}
  \end{subfigure}
  \caption{Available pre-computation example - Text hider}
  \label{fig:painted}
\end{figure}

Datastructure and functions to store, compute, measure, etc. in an unified way are available in utility classes. The execution handler outputs three major elements : 
\begin{itemize}[noitemsep]
\item A \textit{statistics} datastructure : including execution time, pre-computation time, matching time  (total and per element), memory consumption, matching quality metrics ...
\item A \textit{graph} datastructure : each picture matched with another one share a common edge. Nodes are input pictures. The generated graph can directly be read and displayed with tool such as \textbf{VisJS-Classificator} \cite{vincent-circlClassificatorPicturesMatching2019}
\item A \textit{configuration} datastructure : copy of the generated configuration file that conducted to this result
\end{itemize} 

\begin{figure}[h!]
  \centering
  \begin{subfigure}[b]{\linewidth}
  	\includegraphics[width=\textwidth]{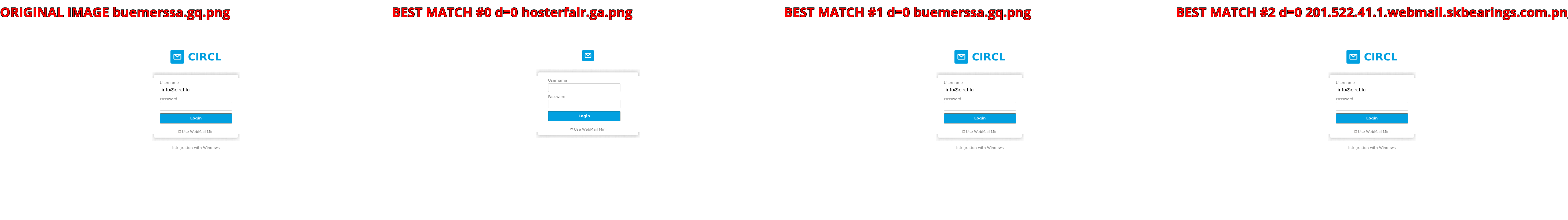}
    \caption{The upper central logo is not matched correctly as the best matching picture (2nd from the left) does not include it}
  \end{subfigure}
  
  \begin{subfigure}[b]{\linewidth}
    \includegraphics[width=\textwidth]{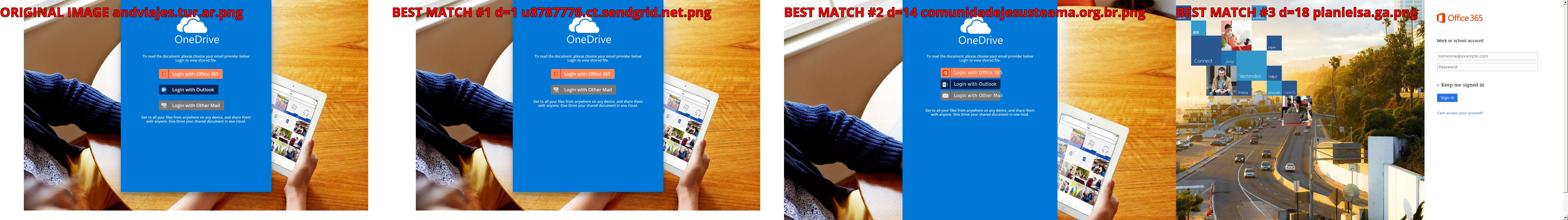} 
    \caption{Login form is matched correctly, as the first matches are similar pictures, even with missing buttons}
  \end{subfigure}
  
  \begin{subfigure}[b]{\linewidth}
    \includegraphics[width=\textwidth]{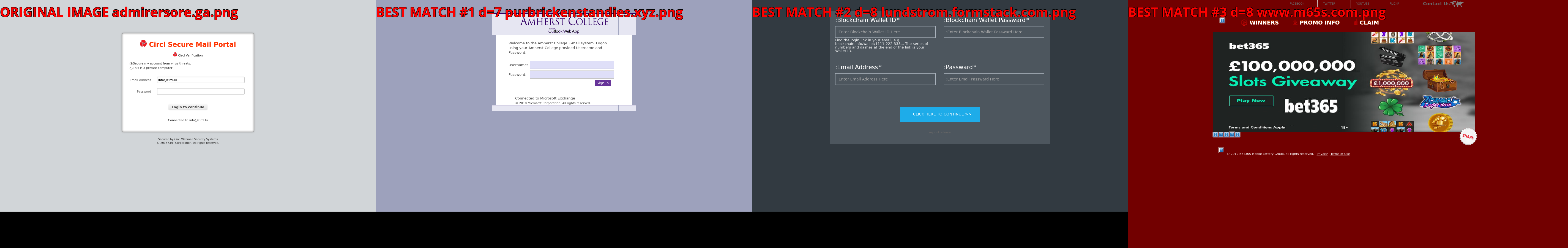} 
    \caption{The form is matched with other forms, but none identical to original one.}
  \end{subfigure}

\caption{Request picture (on the left) and top matches (on the right, top 1 to top 3 from left to right).}
\label{fig:mismatch_logo}
\end{figure}

One can activate the picture exportation of each pair of pictures matched together. This option is not activated by default, due to the high disk-space and computation time (mainly I/O access). However, this output is really relevant to manually spot drawbacks of algorithms. Figure \ref{fig:mismatch_logo} is an example of such output.

After all configuration being evaluated, a wrapper performs a reporting about executed experiments by merging \textit{per-configuration} results (execution time, quality ..) in unified views. These views are: 
\begin{itemize}[noitemsep]
\item An \textit{overview} : a list of configuration name and quality measures, sorted by quality
\item A \textit{\LaTeX\ overview} : same list as previous one, formatted as a \LaTeX -ready table
\item An \textit{inclusion} matrix : a matrix displays the inclusion factor between output graphs. This allows to spot which algorithm is similar to which other algorithm (strictly, its output is included). An instance of the intersection matrix is provided in Appendix \ref{app:A}.
\item A \textit{quality per pair} matrix : by pairing algorithms output per 2 and evaluating the merged graph, it generates a quick overview of which algorithms pairing is wise
\end{itemize} 

Please note that Figure \ref{fig:image_matching_pipeline} presents only an high level view of the framework. Implementation is more precisely described in Annex \ref{app:view_framework}.

\section{Results}
\subsection{Used libraries}
In the following, we're going to present few \textit{relevant visualisation} of the framework outputs, as well as \textit{overview} results as described previously.

Few Image matching libraries were tested, including : 
\begin{itemize}[noitemsep]
\item \textbf{ImageHash}\footnote{\href{https://github.com/JohannesBuchner/imagehash}{github.com/JohannesBuchner/imagehash}} which includes a wide list of fuzzy hash algorithm such as AHash, DHash, PHash, WHash, ... The purpose of these algorithms is to map input files (here, images) into a limited hash space, such that two "similar" images have similar hashes. Their objective is therefore distinct from traditional hash algorithms, which seek to maximize the difference between hashes for even a minimal difference (ultimately, 1 bit) in the input file.
\item \textbf{TLSH}\footnote{\href{https://github.com/trendmicro/tlsh}{github.com/trendmicro/tlsh}}\cite{oliverTLSHLocalitySensitive2013} is also a fuzzy hashing algorithm, available in its own library.
\item \textbf{OpenCV}\footnote{\href{https://github.com/opencv/opencv}{github.com/opencv/opencv}} is an open source computer vision and machine learning software library, which provides a common infrastructure for computer vision applications. Relevant algorithms are available within it, such as SIFT \cite{loweDistinctiveImageFeatures2004}\cite{oteroAnatomySIFTMethod2014} (Scale-invariant Feature Transform), SURF \cite{baySURFSpeededRobust2006} (Speeded Up Robust Features), ORB \cite{rubleeORBEfficientAlternative2011} (Oriented FAST and Rotated BRIEF) ... These algorithms use key point descriptors, a way to condense distributed and locally relevant information, into a vector or set of vectors. These vectors can then be compared.
 However, we focused on open-source implementations and avoided patented implementation. SIFT and SURF were therefore outside of the scope. Thanks to a large performance overview \cite{tareenComparativeAnalysisSIFT2018} we chose to primarily focus on ORB.
\end{itemize} 

Few tips to analyze ORB-matching pictures : 
\begin{itemize}[noitemsep]
\item
\textbf{Parrallel lines} (if there is not rotation) are indicators of quality matching. It keeps the spatial consistency between source and candidate pictures.
\item 
\textbf{Text} seems to be a problem. Letters are matched to letters, generating false positive. It also "uses" descriptor space (number of descriptors is artificially limited), and so, hinders true logo (for example) to be described and used.
\end{itemize}

Graph visualization is valuable to spot unexpected matches or mismatches and quickly navigate through all pictures \cite{vesseyCognitiveFitEmpirical1991a}. Even if this kind of visualization is hard to scale up to billion of pictures, it is still a relevant tool.

\subsection{Examples}
\subsubsection{Fuzzy Hash and ORB}
\begin{figure}[h!]
\centering  

  \begin{subfigure}[b]{0.49\linewidth}
    \includegraphics[width=\linewidth]{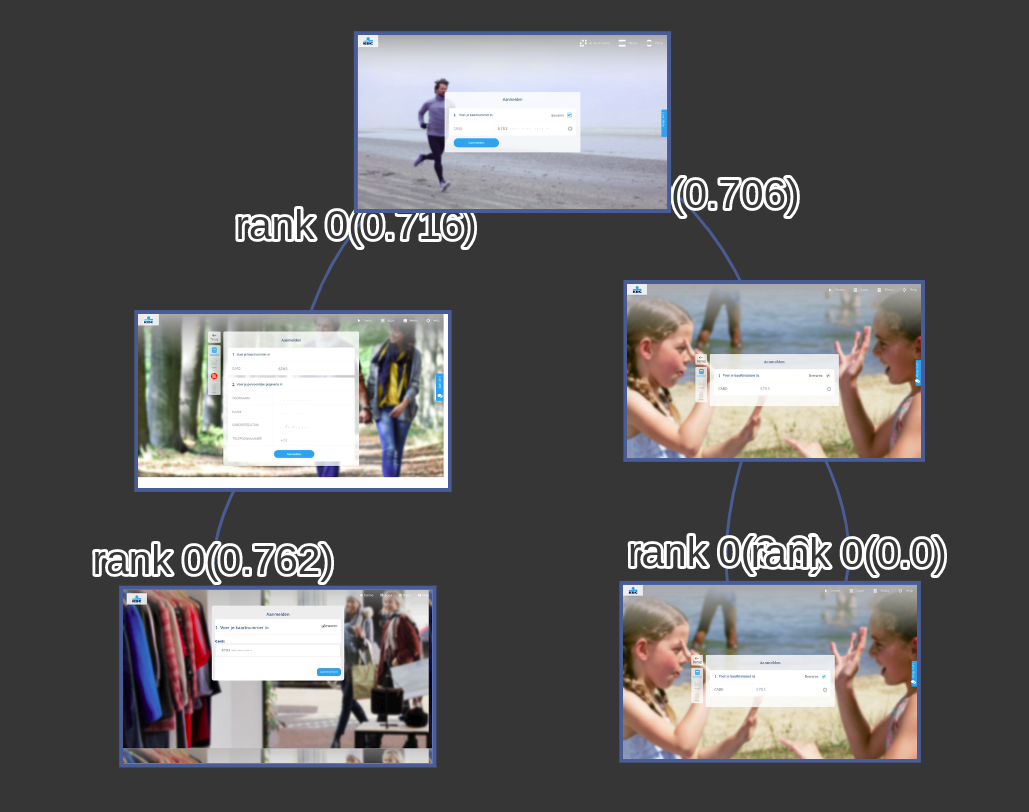}
    \caption{Good brand matching even with different backgrounds. Logo is present on each picture in upper-left corner  - ORB}
    \label{fig:nice_match}
  \end{subfigure}
  \hfill
  \begin{subfigure}[b]{0.4\linewidth}
    \includegraphics[width=\linewidth]{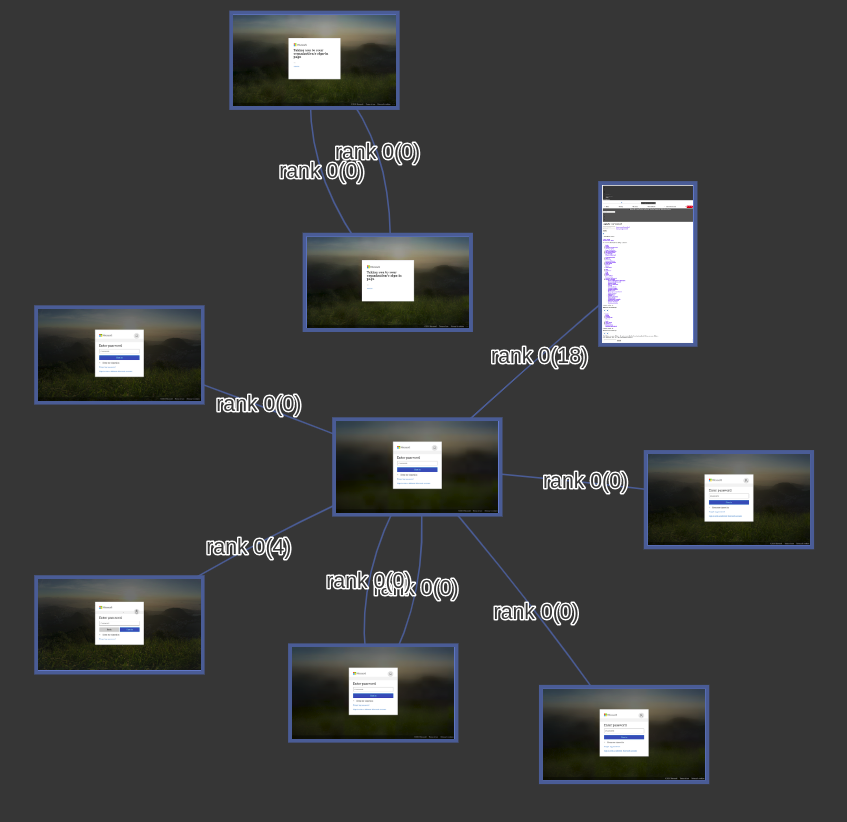}
    \caption{Graph visualisation of \textit{Microsoft} phishing websites - WHash}
    \label{fig:microsoft_mismatch}
  \end{subfigure}

  \begin{subfigure}[b]{0.7\linewidth}
  	\includegraphics[width=\textwidth]{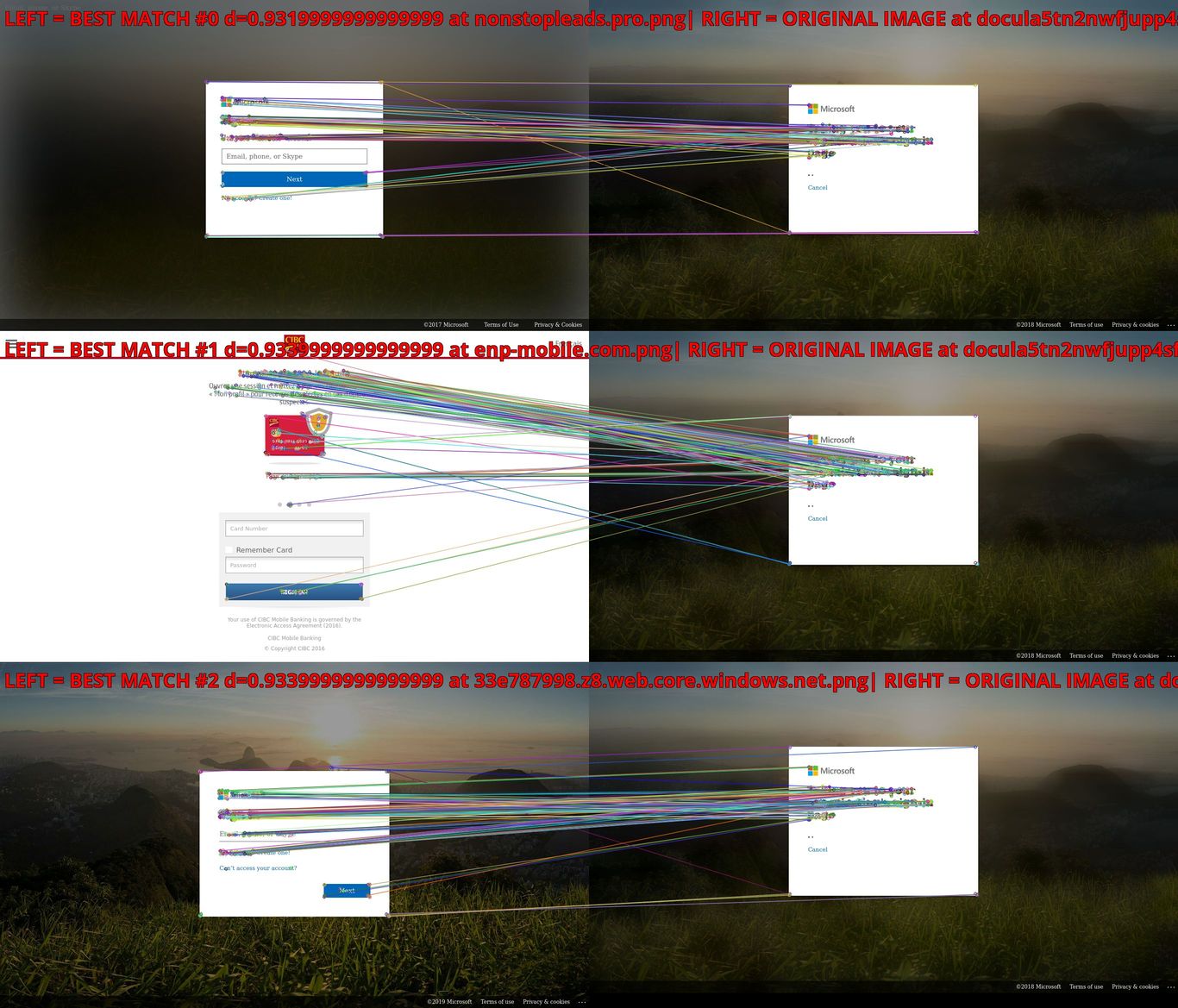}
    \caption{Matches lines between text and forms' corners}
  \end{subfigure}
  
  \caption{Graph visualization and successes}
  \label{goodmatches}
\end{figure}

Two examples are presented. Figure \ref{fig:microsoft_mismatch} presents a mainly correct matching graph including one mismatching edge. This means that the library (here, W-Hash from ImageHash) had correctly detected Microsoft forms as being the sames, but incorrectly gave a Microsoft form as being the closest picture of the white webpage on the right.

Figure \ref{fig:nice_match} presents a match on logo, between globally very different pictures. This means that even if the global look of pictures is different, the algorithm (here, ORB from OpenCV) correctly matched common parts of these screenshots (forms, logo in the corner ...)

\begin{figure}[h!]
  \centering

	\begin{subfigure}[b]{\textwidth} \centering 
	\includegraphics[width=\textwidth]{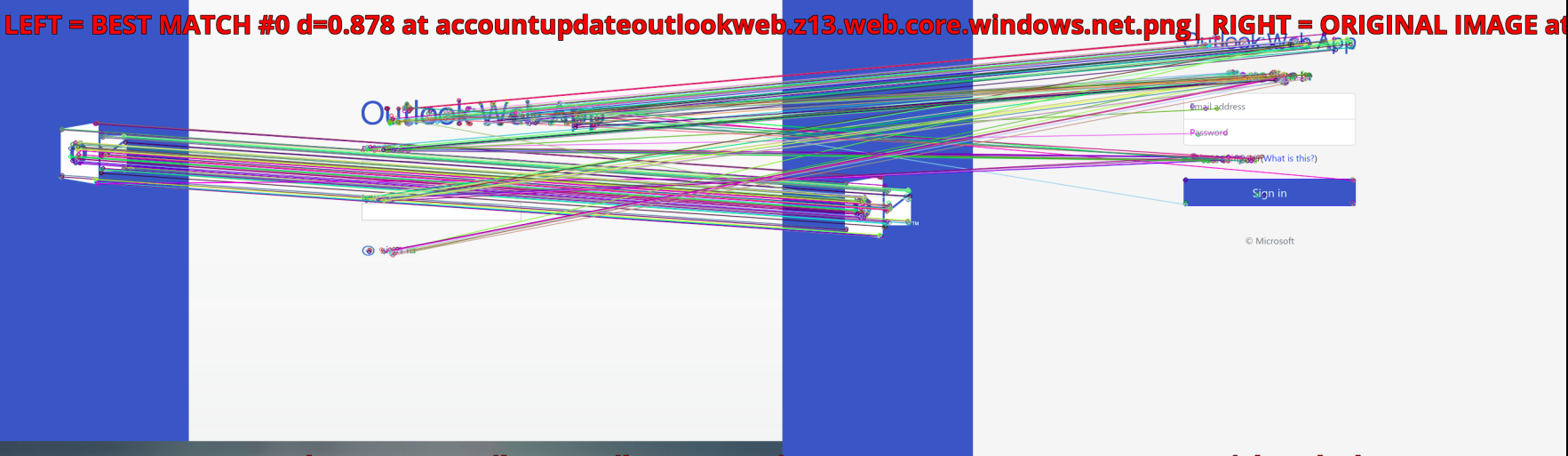} 
	\caption{Good matching on brand even with disturbed background} \end{subfigure}

  	\begin{subfigure}[b]{\linewidth}
    \includegraphics[width=\linewidth]{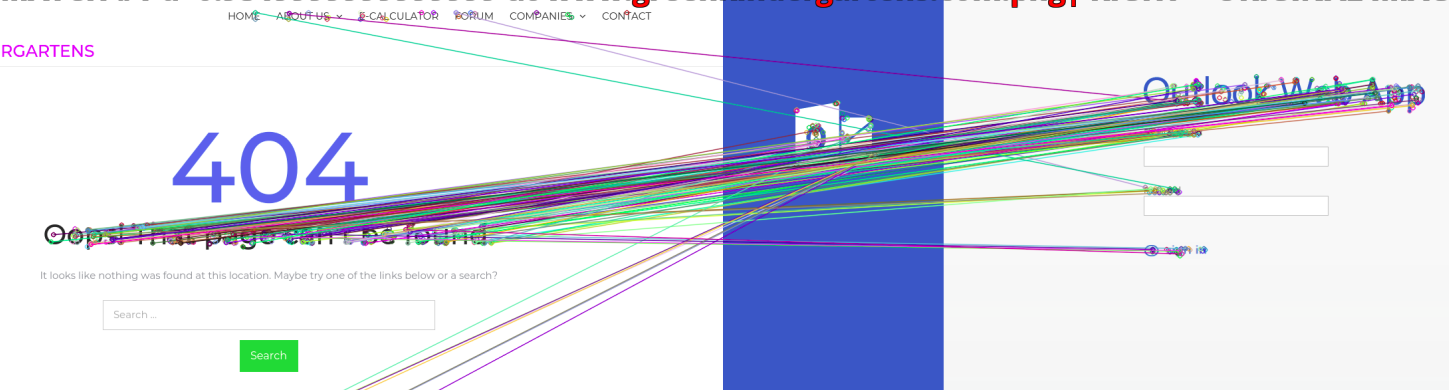}
    \caption{Mismatch due to text/font matching - ORB}
  	\end{subfigure}
  
	\begin{subfigure}[b]{\textwidth} \centering 
	\includegraphics[width=\textwidth]{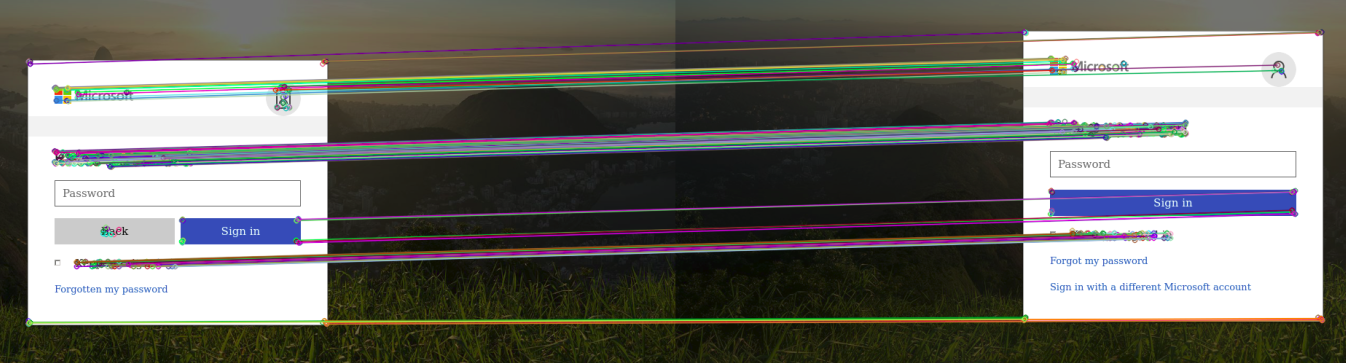} 
	\caption{Perfect match on Microsoft form} \end{subfigure}

  \caption{Graph visualization and issues}
  \label{fig:painted}
\end{figure}

\clearpage

\clearpage
\subsubsection{ORB and RANSAC}

\begin{figure}[h!]

 \centering 
\begin{subfigure}[b]{0.45\textwidth} \centering 
\includegraphics[width=\textwidth]{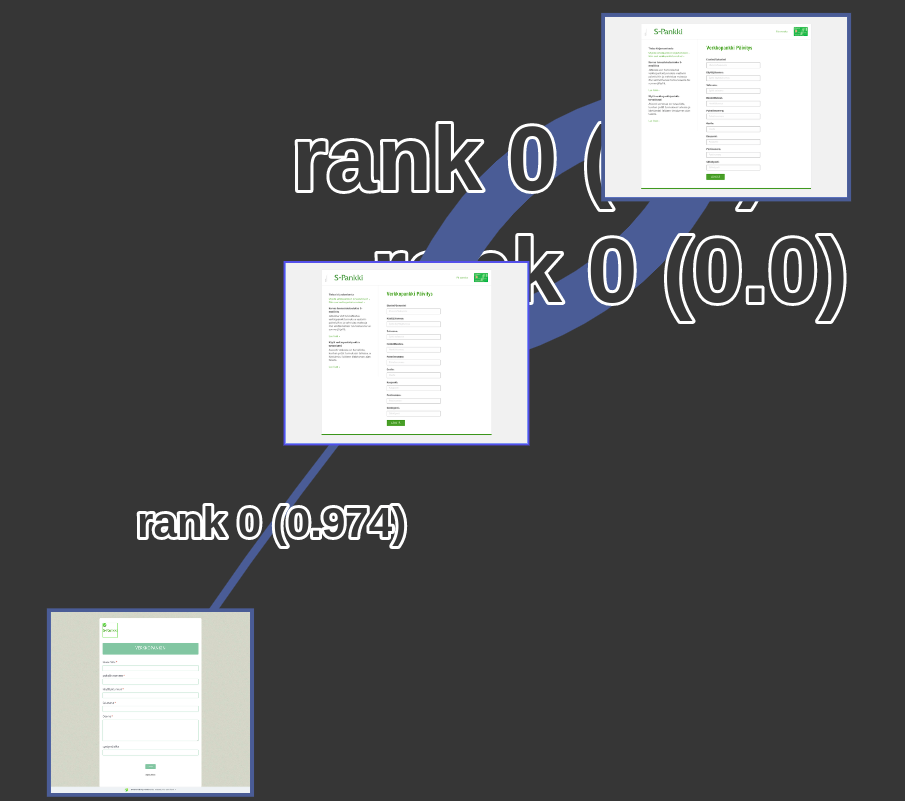} 
\caption{Good match over logo. However one match is above the "critical threshold" of 0.96} \end{subfigure}
\hfill
\begin{subfigure}[b]{0.54\textwidth} \centering 
\includegraphics[width=\textwidth]{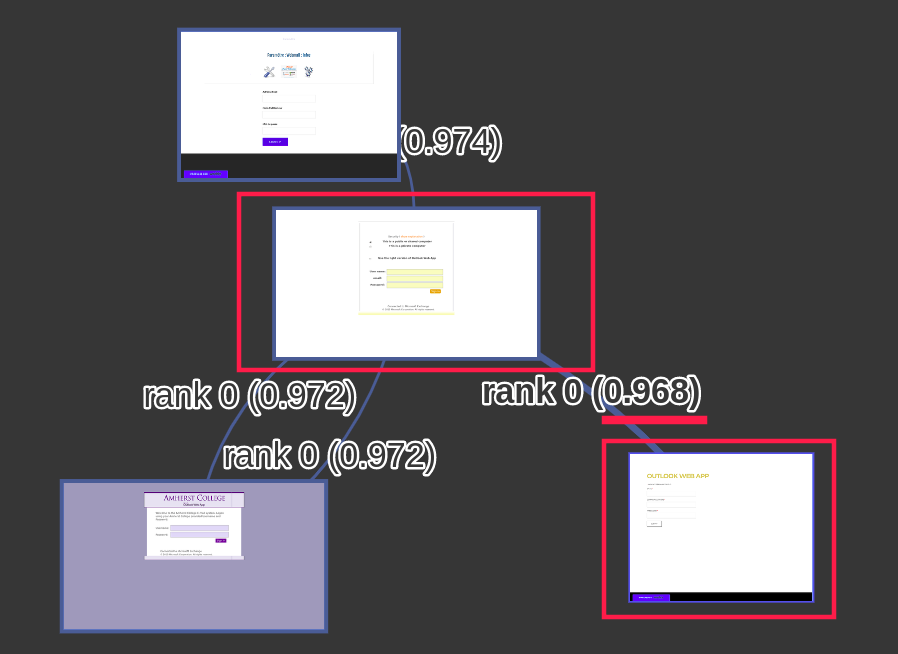} 
\caption{Good match for Outlook, but also above the threshold.} \end{subfigure}

 \centering 
\begin{subfigure}[b]{0.5\textwidth} \centering 
\includegraphics[width=\textwidth]{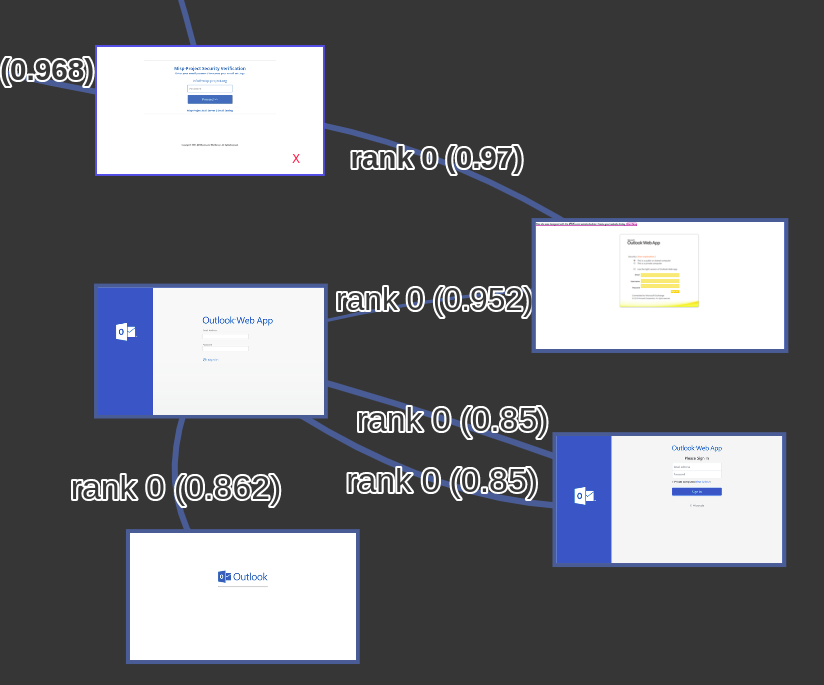} 
\caption{Cluster of Outlook pictures, separated to others by a link above the 0.97 threshold.} \end{subfigure}
\hfill
\begin{subfigure}[b]{0.48\textwidth} \centering 
\includegraphics[width=\textwidth]{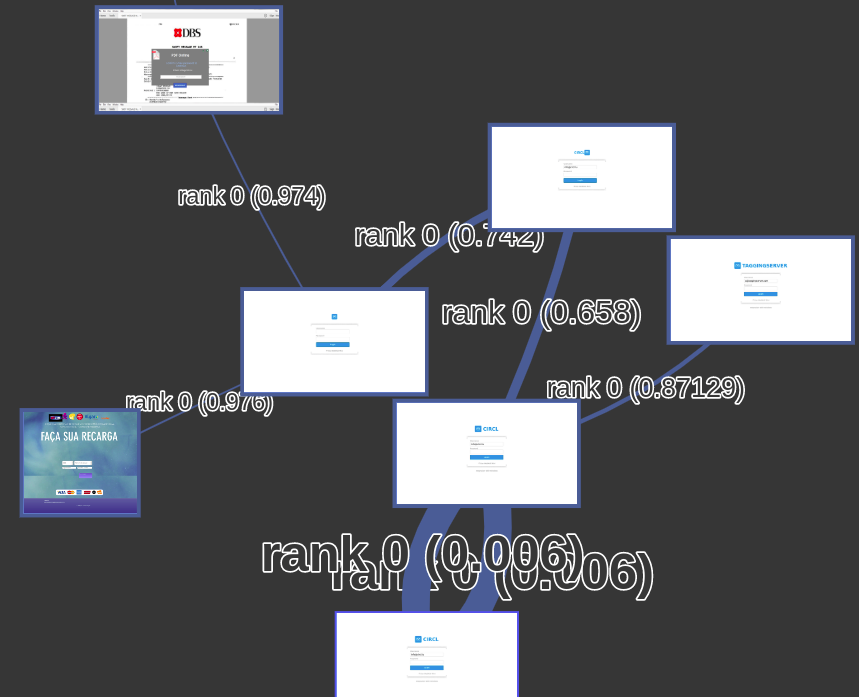} 
\caption{Good cluster of simple blue forms} \end{subfigure}

\caption{Results - ORB - RANSAC Filtering - No matrix filter} 
 \label{fig:ransac1}
\end{figure}

\textbf{RANSAC} outputs a homography between two compared pictures. RANSAC ouputs a homography with a transformation matrix, which also determines which matches are insiders and which are outliers. A 'strong transformation' is a significant rotation/translation/scale-up or down/deformation. A 'light transformation' is a near direct translation, without rotation, scaling or deformation. Figure \ref{fig:ransac1} and \ref{fig:lightreansfo} are examples.

\begin{figure}[h!] 
 \centering
\begin{subfigure}[b]{0.9\textwidth} \centering 
\includegraphics[width=\textwidth]{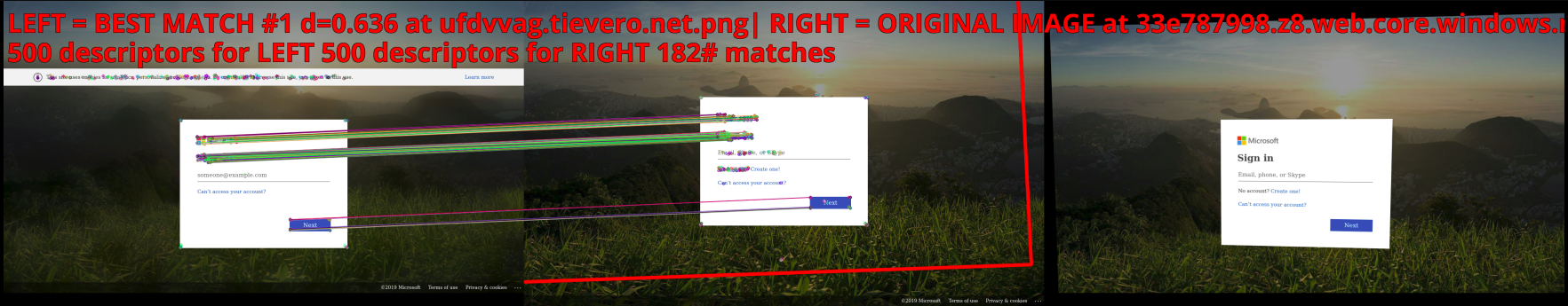} 
\caption{Near no transformation needed. Should be detected as good match.} \label{lightreansfo} \end{subfigure}

 \centering 
\begin{subfigure}[b]{0.9\textwidth} \centering 
\includegraphics[width=\textwidth]{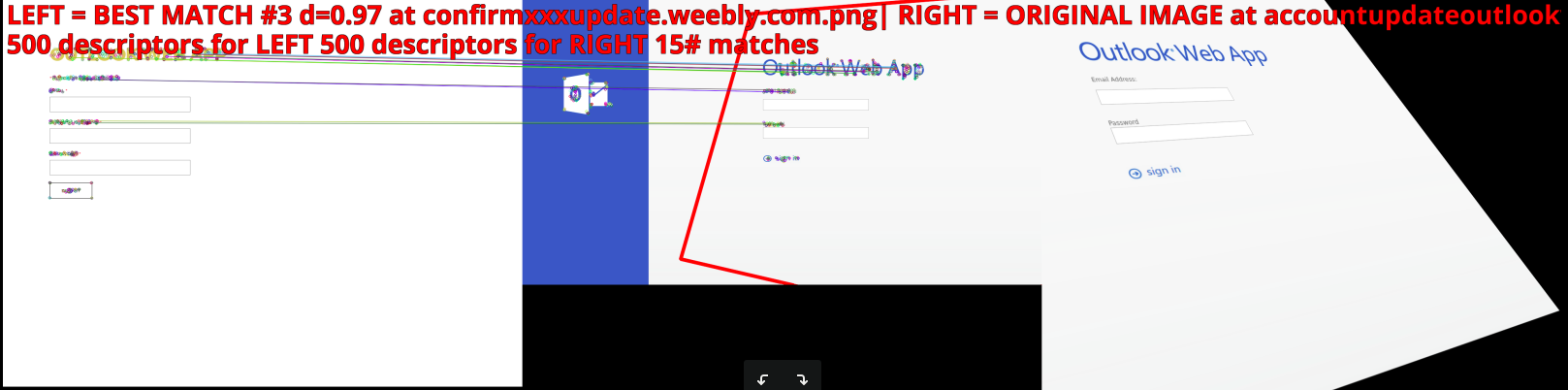} 
\caption{Light distortion. Should be detected as good match.} \end{subfigure}

 \centering 
\begin{subfigure}[b]{0.9\textwidth} \centering 
\includegraphics[width=\textwidth]{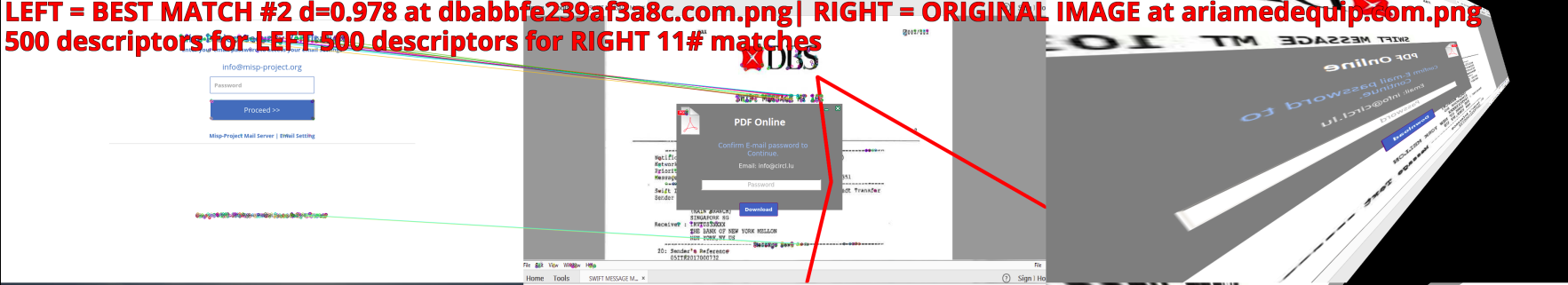} 
\caption{High distortion. Could be accepted as a false positive.} \end{subfigure}

 \centering 
\begin{subfigure}[b]{0.9\textwidth} \centering 
\includegraphics[width=\textwidth]{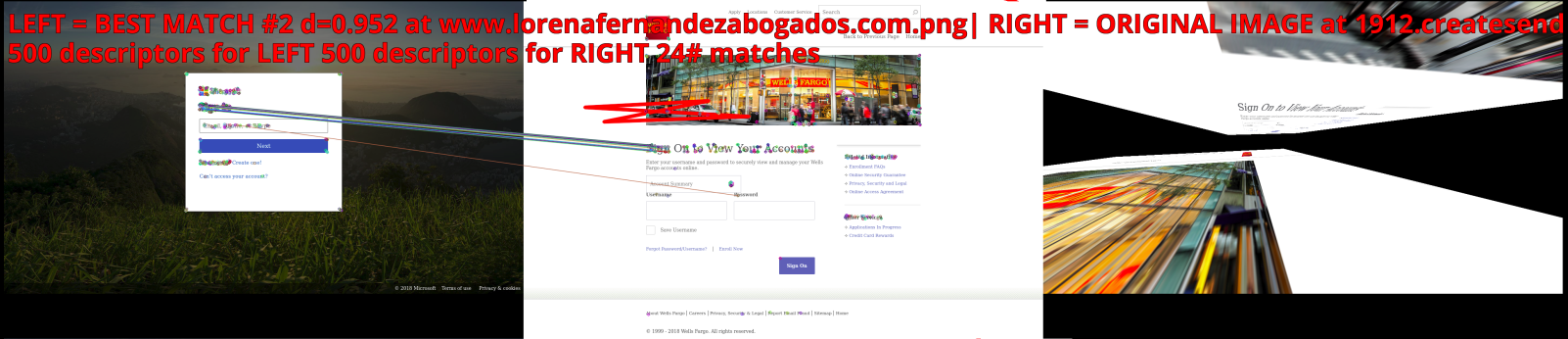} 
\caption{Butterfly pattern with a very heavy transformation. Should be a negative match. Note that the distance score (0.95) would have been very under the expected threshold of 0.96 and therefore not discarded without matrix analysis.} \end{subfigure}

 \centering 
\begin{subfigure}[b]{0.9\textwidth} \centering 
\includegraphics[width=\textwidth]{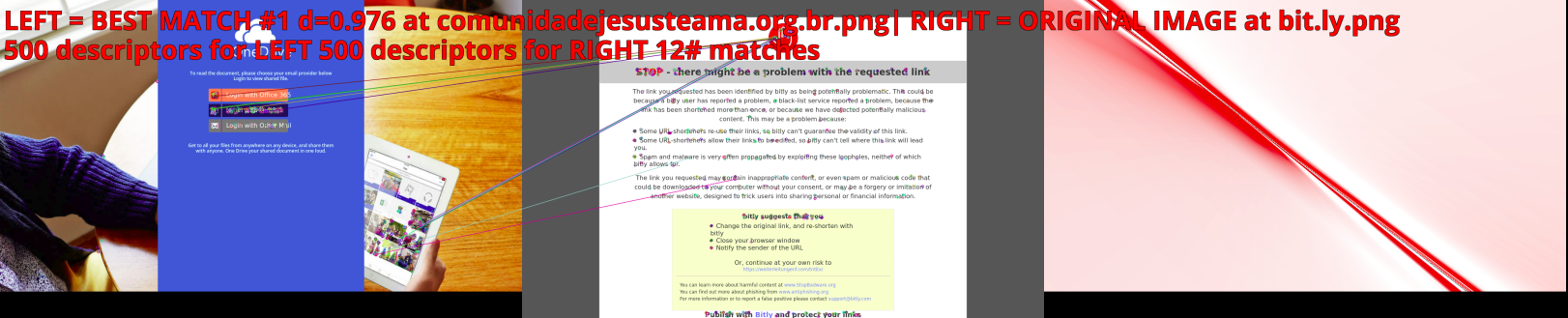} 
\caption{Very high distortion, should be discarded.} \end{subfigure}

\centering
\caption{Matrix transformation visualisation - ORB - RANSAC Filtering - Visualisation of transformation matrix applied to request picture. From left to right : database picture (example), target picture (request), deformed target picture thanks to RANSAC transformation matrix } 
\label{fig:matrixtransformation}

\end{figure}

The transformation matrix applied to the first picture generates a new picture that "fits" the second picture it is compared to. Displaying the first picture with its transformation gives an idea of "how much" the first picture should be transformed. If the transformation is strong (high distortion) then the match is probably low.
Examples are presented in Figure \ref{fig:matrixtransformation}. Please note that the usual name for this operation is "Reprojection error".

Even if an Homography is an interesting tool, it can provides a very wide range of result. Transformation can be "infinite" and so results to false matches. On the contrary, in a real-world context, object can have perspective.
 An affine transformation could be more relevant in our case, as in screenshots, the "depth" won't be used.

\begin{figure}[h!]
  \centering
  \begin{subfigure}[b]{\linewidth}
    \includegraphics[width= \textwidth]{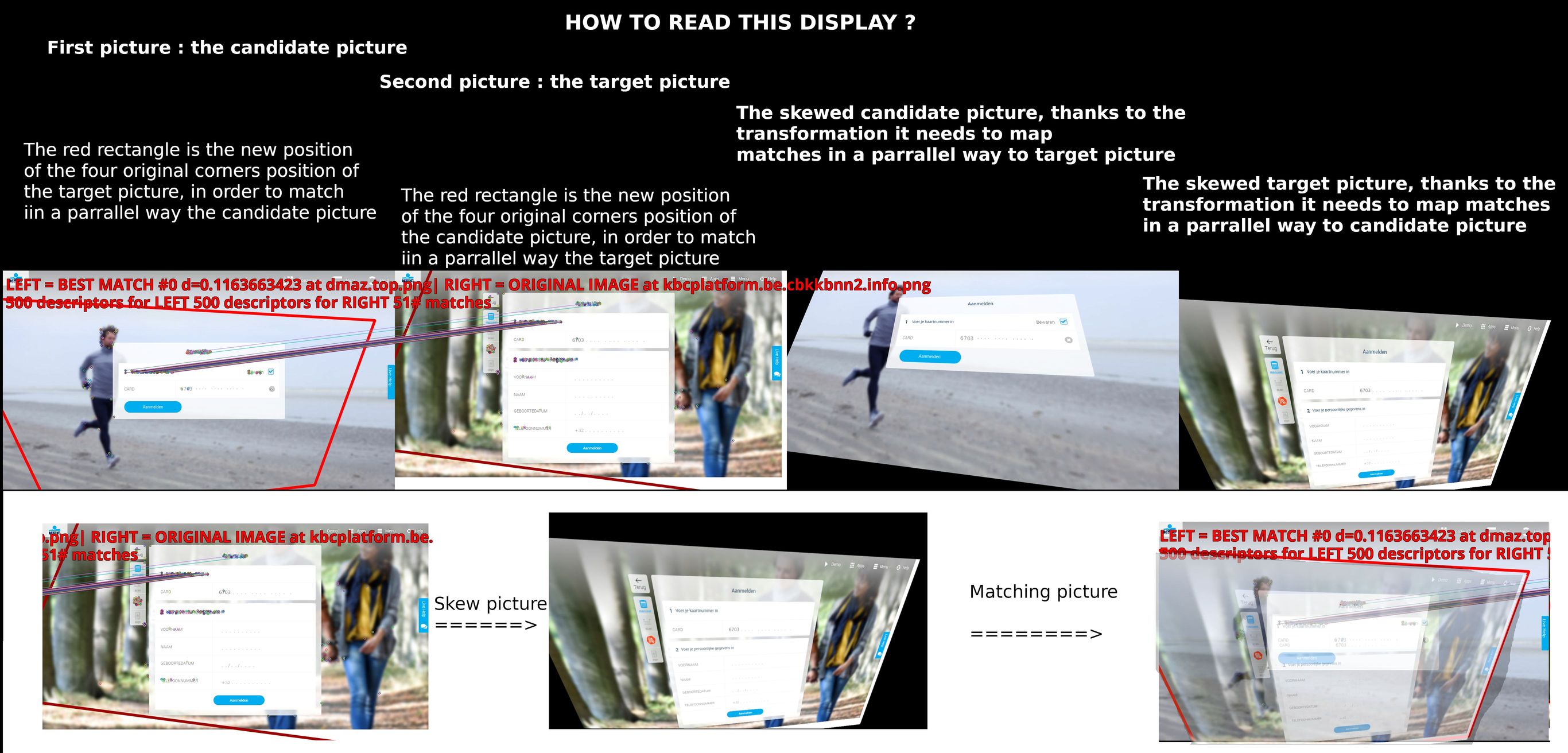} 
    \caption{Explanation of RANSAC visualisation created by the test framework}
  \end{subfigure}  
  \caption{ORB and RANSAC algorithm}
\end{figure}

\begin{figure}[h!]
  \centering
  \begin{subfigure}[b]{\linewidth}
    \includegraphics[width=\textwidth]{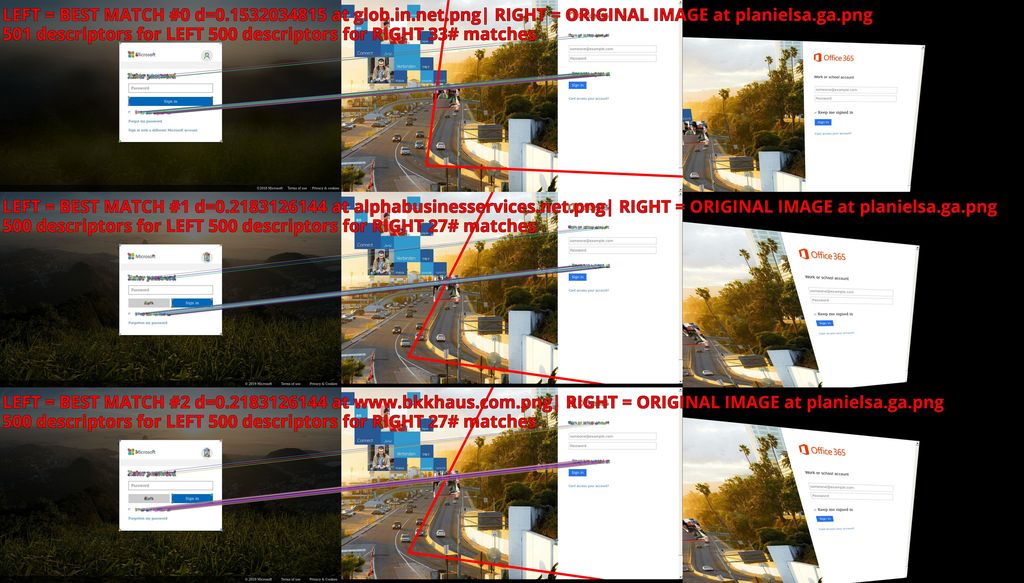} 
    \caption{Match on the form and the relative space between its components}
  \end{subfigure}
      
  \begin{subfigure}[b]{\linewidth}
    \includegraphics[width=\textwidth]{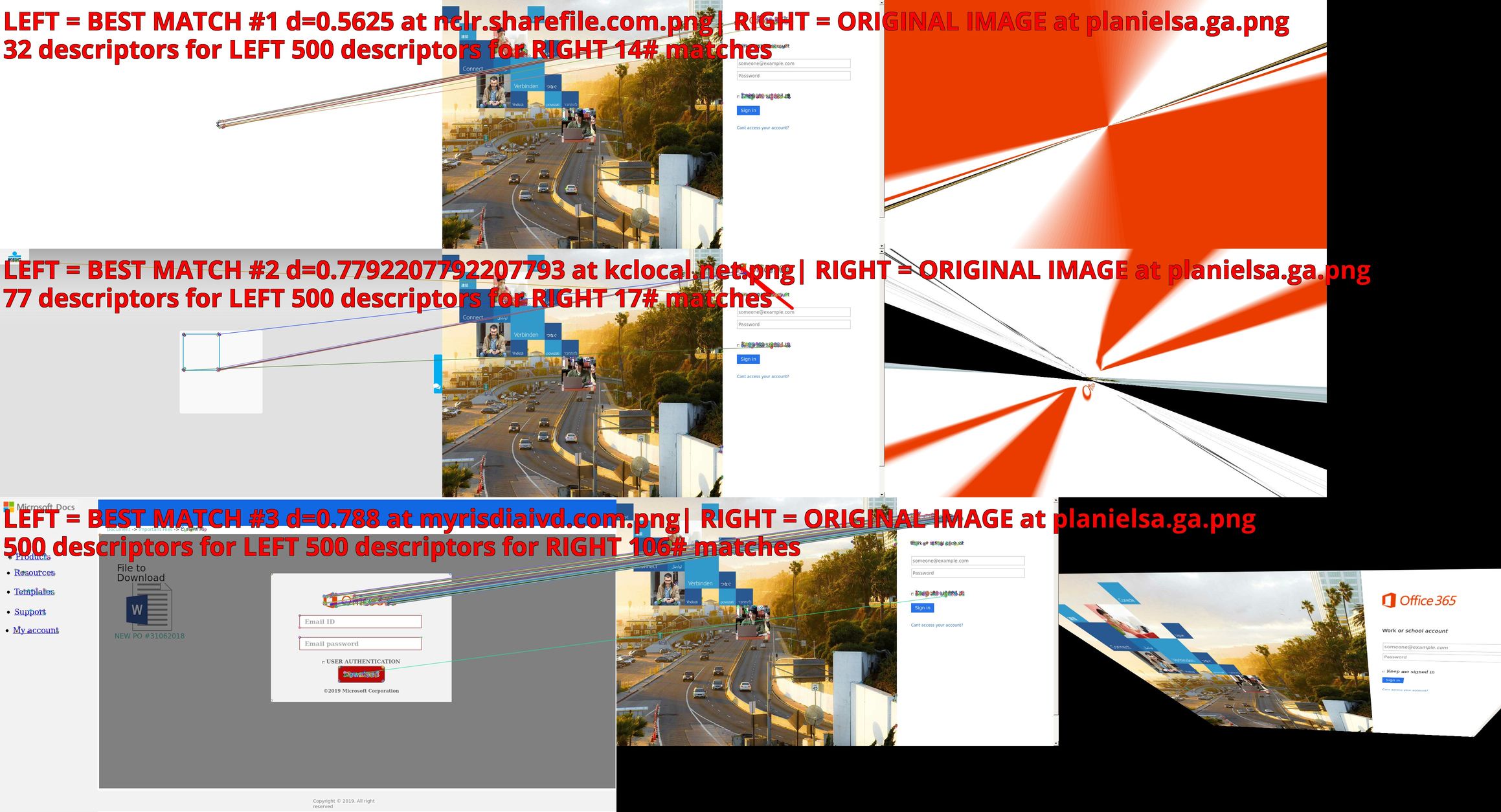} 
    \caption{Strong deformation for the first two matches. Low deformation for the third row.}
  \end{subfigure}
  
  \caption{The first column is the candidate picture, second column is the request picture, third column is the transformed request picture "to match the candidate picture". Less stretched is the best.}
\end{figure}

A few other matches are displayed in Figure \ref{fig:ransac1}. These matches were challenging : various range of colors between pictures of the same category, various backgrounds, etc.

\clearpage
\section{Challenges}

We drawn a few conclusions and encountered a few problems, that need to be discussed for further improvements:
\begin{itemize}[noitemsep]
\item \textbf{ORB can be very sensitive} to input parameters. Fuzzy-hash algorithms have few parameters and so give consistent but untweakable results;
\item \textbf{ORB outperforms} fuzzy-hash algorithms in almost all sounds configuration (about 10\% more true positive matches); 
\item Even if some fuzzy hashes algorithms should not be sensitive to input format (due to internal conversion), it seems that input format generate difference in output results. Uncompressed format (e.g. BMP) gives best results; 
\item Fuzzy Hashs algorithms are quite \textbf{insensitive to text and high frequency} noise. Even if this fact seems to be obvious, it is confirmed by the benchmark;
\item \textbf{Text noise} is a huge issue to feature-matching. This usually does not constitute a problem on natural-scenes used to evaluate these algorithms such as SIFT\cite{loweDistinctiveImageFeatures2004}\cite{oteroAnatomySIFTMethod2014}, SURF\cite{baySURFSpeededRobust2006}, ORB\cite{rubleeORBEfficientAlternative2011}\cite{ORBOrientedFAST2014}, ... However text is prominent in screenshots. This leads to mismatch and unreasonable results. See Figure \ref{fig:404match}. ORB is highly \textbf{sensitive to text}, as corners detectors are detecting corners in each letter. This drawback generates false positive matches and so less true positive matches, in our settings. Hiding the text (blur boxes, mean color boxes, black or white color boxes, ... detected by OCR or DeepLearning approaches) does not fundamentally changes the quality of ORB detection. OCR and Text-hider preprocessing can partially prevent this issue. It seems that the increment in performance is bounded by the imperfectness of OCR used to detected the text to hide. The main challenge for the text-hider is to remove only irrelevant text, while keeping logos and distinctive typography. OCR can detect logos as text, and so hides it, leading to less information for ORB to correctly match pictures.
Please note that no matching had been done on extracted text.
\item Speed-wise and Memory-wise, ORB needs order of magnitude more resources ($10^{-3}$ seconds for hashes, $10^0$ seconds for ORB) than fuzzy-hashs algorithms. This gap is greatly reduced, without a drop of ORB performances, by using Bag-Of-Word/Bag-of-Pictures approaches. Instead of performing $O(N^2)$ comparison per pair of picture (N being the number of descriptor per picture), we can only perform the comparison in constant time, by boiling down N descriptors in a boolean array representing few chosen descriptor presences.
\item RANSAC filtering - verifying homography quality between two matched pictures, which boils down to "Does the picture A needs to be stretched a lot to match picture B ?" - provides the best results of the benchmark, but cost the most ($10^{1}$ seconds for ORB + RANSAC).
\item \textbf{Scalability} : Hash-based comparison are order of magnitude faster than feature-matching algorithm. Even if our current implementation is in a test-state - where preformance is not a main objective - scalability constitutes a lurking and rising issue.
Trying to get the closest picture of each one of a 5000 pictures dataset took 10H with the most "advanced" configuration (ORB+RANSAC based) on a 32 cores, 120Gb RAM server.
\item \textbf{Base model} : Image matching or Image classification? While establish correlation between pictures is the main goal, both approach seems to be able to satisfy it. However, solutions and approaches seems radically different. The final library may use an hybrid approach.
\item \textbf{Algorithm combination and scoring} : Merging algorithms results is a great problem. Their score range is diverse, and even if normalization is doable, the meaning of the score is different. One algorithm can provide a reliable matching if the score is below 0.2 and another one if the score is below 0.6. Therefore, merging results is not only a mathematical issue. 
\item \textbf{Unreachable funnel approach} : A \textit{funnel of algorithm} approach is not adapted to our objective, given current results. This means that using most scalable algorithms first and most-computing-intensive algorithms at the end, while discarding irrelevant candidate picture during the progression in the funnel, is not usable. Correlation between algorithms output is not prevalent and so, what is discarded by one could be matched by another one.
\end{itemize} 

\section{Future work}

Problems presented previously lead to a list of future possible developments : 
\begin{itemize}[noitemsep]
\item Extending provided datasets to support research effort
\item Combining algorithms outputs in an unified and relevant way, to leverage each algorithm main strength
\item Increasing the list of evaluated algorithms, to leverage new main strength
\item Create a challenging dataset with scale/blur/negative/color-changed/.. screenshoots-like pictures, to evaluate the quality of our algorithms combination on edge-cases
\item Evaluate and improve the scalability of the framework, to faster evaluate wider datasets
\item Add more pre and post computation : filter out image, merge image distance, modified edge detector, spatial clustering of descriptors ... to understand what can be done to improve matching quality
\item Time and Memory performance evaluation regarding size of the dataset, to anticipates calculation needs and at which dataset size the framework hits the "complexity wall"
\end{itemize}

\section{Conclusion}
\subsection{Summary}

This research paper proposes that even partial automation of screenshots classification would reduce the burden on security teams.

The presented evaluation framework is available online at \href{https://github.com/CIRCL/carl-hauser}{github.com/CIRCL/carl-hauser}.

Research results were presented along with encountered issues. Algorithms modification are still being performed to solve met issues and improve overall quality of the future library. The issues and their solution presented in this paper constitute a starting point on the path to the creation of a matching and clustering library, intended to be used in Open Source tools.\\
In the context of this paper, these results represents above all a usage example of the provided datasets.

\subsection{Contact information}
\label{contactinfo}
If you have a complaint related to the dataset or the processing over it, please contact us. We aim to be transparent, not only about how we process but also about rights that are linked to such information and processing. 
You can contact us at \href{https://circl.lu/contact/}{circl.lu/contact/} for request about the dataset itself, regarding elements of the dataset, or extension requests.
You can contact us at same address or on \href{https://github.com/CIRCL/carl-hauser}{github}  for feedback about the benchmarking framework, methodology or relevant ideas/inquiries.

\clearpage
\pagebreak
\bibliographystyle{paper-ressources/IEEEtran}
\bibliography{./carl-hauser.bib}

\clearpage
\pagebreak
\part{Appendices}

\section{Detailed view of the framework}
\label{app:view_framework}

The exact implementation differs on some points with the idealized overview presented in the paper body. For example, \textit{core computation handler} and \textit{execution handler} is one element, which can entirely be overwritten. This allows a more flexible and adaptable framework to new algorithms, as each one may need to overwrite larger part of the execution handler.

\begin{sidewaysfigure}[h!]
  \centering
  \includegraphics[width=\textwidth]{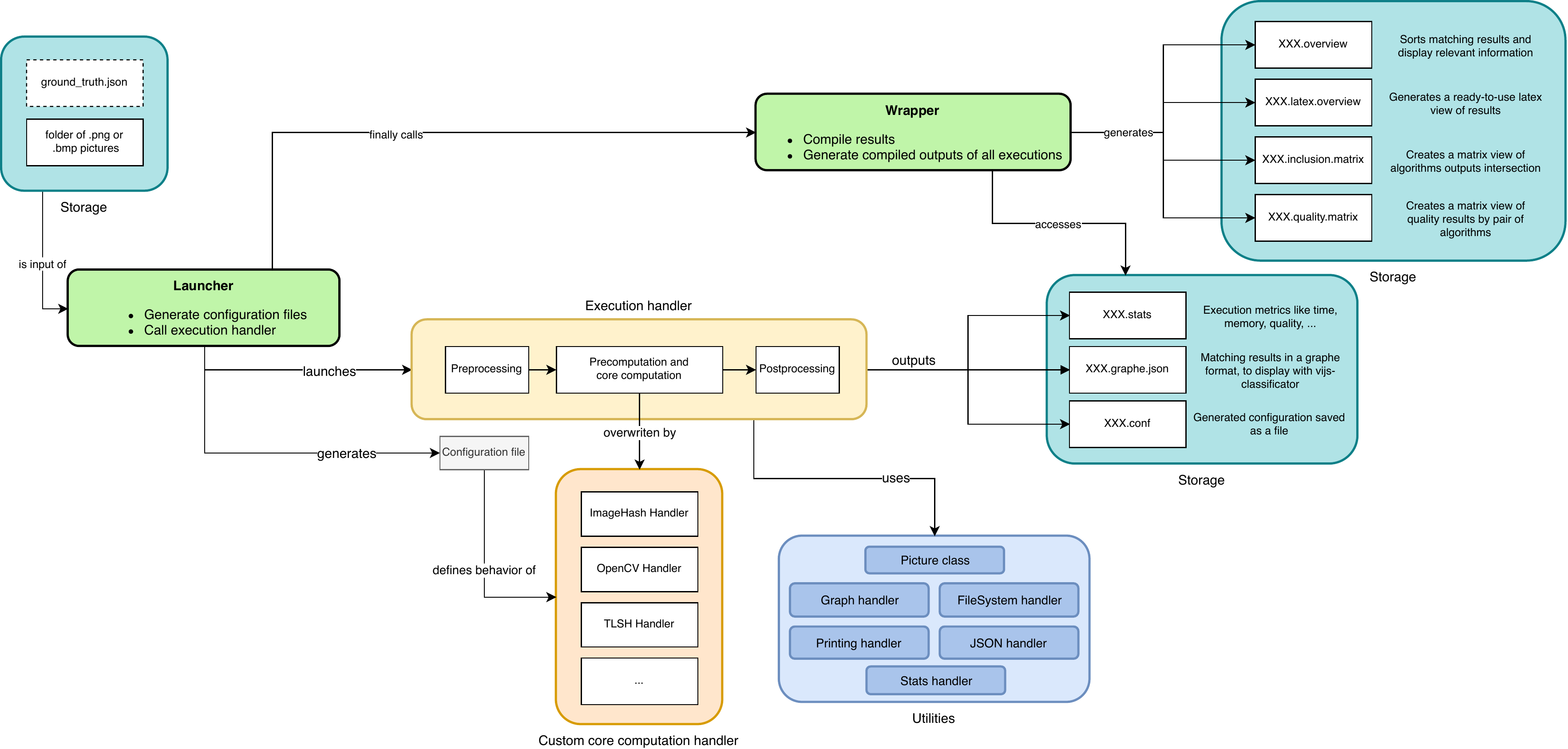}
  \caption{Global overview of test framework}
  \label{fig:image_matching_pipeline_detailed}
\end{sidewaysfigure}

\clearpage
\pagebreak
\section{Result table}

Following table present raw results reduced to best algorithms.

A shortened results table is provided, with algorithms evaluated as "\textit{best ones}" on the \textit{phishing dataset}. This dataset was composed of 207 screenshots of phishing websites. The pre-computation time is the hashing or feature-description time. The matching time is the elapsed time between a request picture and a best-match answer from the algorithm. As explained in previously in the paper body, the true positive rate is the inclusion rate between the output graph of the algorithm and the ground truth graph.

Values are normalized thanks to the maximum obtainable score for this version of the score evaluation. Evaluated algorithms always give a match for a given image, even for outliers, which has no outbound link in the ground truth graph. So even in the best case the algorithm has a bounded maximum score.

\begin{adjustbox}{width=\textwidth} 
\renewcommand{\arraystretch}{1.2}
\label{table:png_scores}
\centering
\begin{tabular}{|c||c|c|c|}

\hline
NAME (PARAMETERS) & TRUE POSITIVE (norm.) & PRE COMPUTING (s) & MATCHING (s)\\ \hline
ORB 500 LEN MAX NO FILTER STD BRUTE FORCE ENA. & 0.65263  (0.77641)& 0.04624 & 1.9087\\ \hline
ORB 500 LEN MIN FAR THR. KNN 2 FLANN LSH ENA. & 0.64211  (0.76389)& 0.04634 & 1.1203\\ \hline
ORB 500 LEN MIN FAR THR. KNN 2 FLANN LSH DIS. & 0.63684  (0.75762)& 0.04614 & 1.12314\\ \hline
ORB 500 LEN MAX FAR THR. KNN 2 FLANN LSH ENA. & 0.63158  (0.75136)& 0.05965 & 1.26475\\ \hline
ORB 500 LEN MAX FAR THR. KNN 2 FLANN LSH DIS. & 0.62632  (0.74511)& 0.04613 & 1.12695\\ \hline
ORB 500 LEN MIN RATIO KNN 2 FLANN LSH DIS. & 0.61053  (0.72632)& 0.04764 & 1.19758\\ \hline
ORB 500 LEN MIN RATIO KNN 2 FLANN LSH ENA. & 0.60526  (0.72005)& 0.04787 & 1.14532\\ \hline
D HASH & 0.60386  (0.71839)& 0.06124 & 0.00214\\ \hline
A HASH & 0.57971  (0.68966)& 0.06268 & 0.00218\\ \hline
D HASH VERTICAL & 0.57488  (0.68391)& 0.05934 & 0.00216\\ \hline
ORB 500 LEN MIN NO FILTER STD BRUTE FORCE ENABLED & 0.56842  (0.67622)& 0.0452 & 2.08476\\ \hline
P HASH & 0.56039  (0.66667)& 0.06403 & 0.00244\\ \hline
W HASH & 0.53623  (0.63793)& 0.1221 & 0.00205\\ \hline
P HASH SIMPLE & 0.52657  (0.62644)& 0.06377 & 0.00222\\ \hline
ORB 500 LEN MAX RATIO BAD STD BRUTE FORCE ENABLED & 0.46316  (0.55101)& 0.0445 & 1.68615\\ \hline
ORB 500 LEN MIN RATIO BAD STD BRUTE FORCE ENABLED & 0.45263  (0.53847)& 0.0445 & 2.15439\\ \hline
TLSH & 0.42512  (0.50575)& 0.00498 & 0.00096\\ \hline
TLSH NO LENGTH & 0.40580 (0.48276)& 0.00494 & 0.0011\\ \hline
\end{tabular}
\end{adjustbox}

\clearpage
\begin{adjustbox}{width=\textwidth}     \renewcommand{\arraystretch}{1.4}

    \begin{tabular}{| p{400pt} || l | l | l |}
        \hline
NAME & TRUE POSITIVE & PRE COMPUTING (s) & MATCHING (s)\\ \hline
raw phishing PNG ORB 500 LEN MAX NO FILTER STD BRUTE FORCE ENABLED & 0.65263 & 0.04624 & 1.9087\\ \hline
raw phishing PNG ORB 500 LEN MIN FAR THREESHOLD KNN 2 BRUTE FORCE DISABLED & 0.64737 & 0.06641 & 1.23615\\ \hline
raw phishing PNG ORB 500 LEN MIN RATIO CORRECT KNN 2 BRUTE FORCE DISABLED & 0.64737 & 0.05256 & 1.17277\\ \hline
raw phishing PNG ORB 500 LEN MAX RATIO CORRECT KNN 2 BRUTE FORCE DISABLED & 0.64737 & 0.04775 & 1.34641\\ \hline
raw phishing PNG ORB 500 LEN MAX FAR THREESHOLD KNN 2 BRUTE FORCE DISABLED & 0.64737 & 0.06434 & 1.20571\\ \hline
raw phishing PNG ORB 500 LEN MIN FAR THREESHOLD KNN 2 FLANN LSH ENABLED & 0.64211 & 0.04634 & 1.1203\\ \hline
raw phishing PNG ORB 500 LEN MIN FAR THREESHOLD KNN 2 FLANN LSH DISABLED & 0.63684 & 0.04614 & 1.12314\\ \hline
raw phishing PNG ORB 500 LEN MAX FAR THREESHOLD KNN 2 FLANN LSH ENABLED & 0.63158 & 0.05965 & 1.26475\\ \hline
raw phishing PNG ORB 500 LEN MAX FAR THREESHOLD KNN 2 FLANN LSH DISABLED & 0.62632 & 0.04613 & 1.12695\\ \hline
raw phishing PNG ORB 500 LEN MIN RATIO CORRECT KNN 2 FLANN LSH DISABLED & 0.61053 & 0.04764 & 1.19758\\ \hline
raw phishing PNG ORB 500 LEN MAX RATIO CORRECT KNN 2 FLANN LSH DISABLED & 0.61053 & 0.04769 & 1.13552\\ \hline
raw phishing PNG ORB 500 LEN MIN RATIO CORRECT KNN 2 FLANN LSH ENABLED & 0.60526 & 0.04787 & 1.14532\\ \hline
raw phishing PNG ORB 500 LEN MAX RATIO CORRECT KNN 2 FLANN LSH ENABLED & 0.60526 & 0.04742 & 1.11723\\ \hline
raw phishing PNG D HASH & 0.60386 & 0.06124 & 0.00214\\ \hline
raw phishing PNG A HASH & 0.57971 & 0.06268 & 0.00218\\ \hline
raw phishing PNG D HASH VERTICAL & 0.57488 & 0.05934 & 0.00216\\ \hline
raw phishing PNG ORB 500 LEN MIN NO FILTER STD BRUTE FORCE ENABLED & 0.56842 & 0.0452 & 2.08476\\ \hline
raw phishing PNG P HASH & 0.56039 & 0.06403 & 0.00244\\ \hline
raw phishing PNG W HASH & 0.53623 & 0.1221 & 0.00205\\ \hline
raw phishing PNG P HASH SIMPLE & 0.52657 & 0.06377 & 0.00222\\ \hline
raw phishing PNG ORB 500 LEN MAX RATIO BAD STD BRUTE FORCE ENABLED & 0.46316 & 0.0445 & 1.68615\\ \hline
raw phishing PNG ORB 500 LEN MIN RATIO BAD STD BRUTE FORCE ENABLED & 0.45263 & 0.0445 & 2.15439\\ \hline
raw phishing PNG TLSH & 0.42512 & 0.00498 & 0.00096\\ \hline
raw phishing PNG TLSH NO LENGTH & 0.4058 & 0.00494 & 0.0011\\ \hline
raw phishing PNG ORB 500 LEN MAX RATIO BAD STD FLANN LSH ENABLED & 0.30526 & 0.05724 & 1.50447\\ \hline
raw phishing PNG ORB 500 LEN MAX RATIO BAD STD BRUTE FORCE DISABLED & 0.3 & 0.07377 & 1.17801\\ \hline
raw phishing PNG ORB 500 LEN MIN RATIO BAD STD BRUTE FORCE DISABLED & 0.28947 & 0.04555 & 0.99388\\ \hline
raw phishing PNG ORB 500 LEN MAX RATIO BAD STD FLANN LSH DISABLED & 0.27368 & 0.14438 & 1.56765\\ \hline
raw phishing PNG ORB 500 LEN MIN RATIO BAD STD FLANN LSH ENABLED & 0.27368 & 0.05961 & 1.47452\\ \hline
raw phishing PNG ORB 500 LEN MIN RATIO BAD STD FLANN LSH DISABLED & 0.26842 & 0.04805 & 1.17483\\ \hline
raw phishing PNG ORB 500 LEN MAX NO FILTER STD FLANN LSH ENABLED & 0.06842 & 0.05417 & 1.1589\\ \hline
raw phishing PNG ORB 500 LEN MAX NO FILTER STD FLANN LSH DISABLED & 0.05789 & 0.04719 & 1.34306\\ \hline
raw phishing PNG ORB 500 LEN MAX NO FILTER STD BRUTE FORCE DISABLED & 0.02105 & 0.0631 & 1.14467\\ \hline
raw phishing PNG ORB 500 LEN MIN NO FILTER STD FLANN LSH DISABLED & 0.00526 & 0.05679 & 1.18021\\ \hline
raw phishing PNG ORB 500 LEN MIN NO FILTER STD FLANN LSH ENABLED & 0.00526 & 0.04684 & 1.08095\\ \hline
raw phishing PNG ORB 500 LEN MIN NO FILTER STD BRUTE FORCE DISABLED & 0.0 & 0.04518 & 0.95656\\ \hline
\end{tabular}

\end{adjustbox}

\begin{adjustbox}{width=\textwidth}     \renewcommand{\arraystretch}{1.4}
    \begin{tabular}{| p{400pt} || l | l | l |}
        \hline
NAME & TRUE POSITIVE & PRE COMPUTING (sec) & MATCHING (sec)\\ \hline
raw phishing bmp BMP ORB 500 LEN MAX NO FILTER STD BRUTE FORCE ENABLED & 0.64737 & 0.07296 & 2.52043\\ \hline
raw phishing bmp BMP D HASH & 0.60386 & 0.02017 & 0.00224\\ \hline
raw phishing bmp BMP TLSH & 0.58937 & 0.08356 & 0.00093\\ \hline
raw phishing bmp BMP A HASH & 0.57971 & 0.02544 & 0.00245\\ \hline
raw phishing bmp BMP TLSH NO LENGTH & 0.57005 & 0.0856 & 0.00114\\ \hline
raw phishing bmp BMP D HASH VERTICAL & 0.56522 & 0.01986 & 0.00228\\ \hline
raw phishing bmp BMP ORB 500 LEN MIN NO FILTER STD BRUTE FORCE ENABLED & 0.55789 & 0.04995 & 2.40043\\ \hline
raw phishing bmp BMP P HASH & 0.55072 & 0.02438 & 0.00232\\ \hline
raw phishing bmp BMP W HASH & 0.54106 & 0.09857 & 0.00221\\ \hline
raw phishing bmp BMP P HASH SIMPLE & 0.52174 & 0.02224 & 0.00233\\ \hline
raw phishing bmp BMP ORB 500 LEN MAX RATIO BAD STD BRUTE FORCE ENABLED & 0.45263 & 0.0637 & 2.45745\\ \hline
raw phishing bmp BMP ORB 500 LEN MIN RATIO BAD STD BRUTE FORCE ENABLED & 0.44211 & 0.06207 & 2.39609\\ \hline
raw phishing bmp BMP ORB 500 LEN MAX RATIO BAD STD BRUTE FORCE DISABLED & 0.28421 & 0.05223 & 1.45008\\ \hline
raw phishing bmp BMP ORB 500 LEN MIN RATIO BAD STD BRUTE FORCE DISABLED & 0.27368 & 0.06659 & 1.3083\\ \hline
raw phishing bmp BMP ORB 500 LEN MAX NO FILTER STD BRUTE FORCE DISABLED & 0.01579 & 0.0531 & 1.43634\\ \hline
raw phishing bmp BMP ORB 500 LEN MIN NO FILTER STD BRUTE FORCE DISABLED & 0.0 & 0.05253 & 1.17698\\ \hline
\end{tabular}
\end{adjustbox}

\begin{adjustbox}{width=\textwidth}         \renewcommand{\arraystretch}{1.4}
    \begin{tabular}{| p{400pt} || l | l | l |}
        \hline
NAME & TRUE POSITIVE & PRE COMPUTING (sec) & MATCHING (sec)\\ \hline
raw phishing COLORED PNG ORB 500 LEN MAX NO FILTER STD BRUTE FORCE ENABLED & 0.59474 & 0.04536 & 1.5378\\ \hline
raw phishing COLORED PNG P HASH & 0.58937 & 0.04627 & 0.00233\\ \hline
raw phishing COLORED PNG D HASH VERTICAL & 0.57488 & 0.04983 & 0.00236\\ \hline
raw phishing COLORED PNG D HASH & 0.57488 & 0.04925 & 0.00301\\ \hline
raw phishing COLORED PNG ORB 500 LEN MIN NO FILTER STD BRUTE FORCE ENABLED & 0.56842 & 0.0739 & 2.11373\\ \hline
raw phishing COLORED PNG ORB 500 LEN MAX RATIO CORRECT KNN 2 BRUTE FORCE DISABLED & 0.55789 & 0.0439 & 0.86356\\ \hline
raw phishing COLORED PNG ORB 500 LEN MAX FAR THREESHOLD KNN 2 BRUTE FORCE DISABLED & 0.55789 & 0.0439 & 0.86493\\ \hline
raw phishing COLORED PNG A HASH & 0.55072 & 0.04406 & 0.00229\\ \hline
raw phishing COLORED PNG ORB 500 LEN MIN RATIO CORRECT KNN 2 BRUTE FORCE DISABLED & 0.54737 & 0.04379 & 0.85381\\ \hline
raw phishing COLORED PNG ORB 500 LEN MIN FAR THREESHOLD KNN 2 BRUTE FORCE DISABLED & 0.54737 & 0.04405 & 0.85917\\ \hline
raw phishing COLORED PNG ORB 500 LEN MAX FAR THREESHOLD KNN 2 FLANN LSH ENABLED & 0.54737 & 0.04403 & 1.08197\\ \hline
raw phishing COLORED PNG ORB 500 LEN MAX FAR THREESHOLD KNN 2 FLANN LSH DISABLED & 0.54211 & 0.04388 & 1.084\\ \hline
raw phishing COLORED PNG ORB 500 LEN MAX RATIO CORRECT KNN 2 FLANN LSH ENABLED & 0.53684 & 0.04413 & 1.08097\\ \hline
raw phishing COLORED PNG ORB 500 LEN MIN FAR THREESHOLD KNN 2 FLANN LSH DISABLED & 0.53684 & 0.0441 & 1.07795\\ \hline
raw phishing COLORED PNG ORB 500 LEN MAX RATIO CORRECT KNN 2 FLANN LSH DISABLED & 0.53158 & 0.0439 & 1.08247\\ \hline
raw phishing COLORED PNG P HASH SIMPLE & 0.52657 & 0.04703 & 0.00298\\ \hline
raw phishing COLORED PNG ORB 500 LEN MIN FAR THREESHOLD KNN 2 FLANN LSH ENABLED & 0.52105 & 0.04423 & 1.08245\\ \hline
raw phishing COLORED PNG W HASH & 0.51208 & 0.11247 & 0.00229\\ \hline
raw phishing COLORED PNG ORB 500 LEN MIN RATIO CORRECT KNN 2 FLANN LSH ENABLED & 0.51053 & 0.04389 & 1.0917\\ \hline
raw phishing COLORED PNG ORB 500 LEN MIN RATIO CORRECT KNN 2 FLANN LSH DISABLED & 0.5 & 0.0438 & 1.08247\\ \hline
raw phishing COLORED PNG TLSH & 0.47826 & 0.00513 & 0.00096\\ \hline
raw phishing COLORED PNG TLSH NO LENGTH & 0.45411 & 0.00508 & 0.00111\\ \hline
raw phishing COLORED PNG ORB 500 LEN MAX RATIO BAD STD BRUTE FORCE ENABLED & 0.44737 & 0.0706 & 2.30236\\ \hline
raw phishing COLORED PNG ORB 500 LEN MIN RATIO BAD STD BRUTE FORCE ENABLED & 0.38947 & 0.0676 & 2.39523\\ \hline
raw phishing COLORED PNG ORB 500 LEN MIN RATIO BAD STD FLANN LSH ENABLED & 0.24211 & 0.04551 & 1.05413\\ \hline
raw phishing COLORED PNG ORB 500 LEN MAX RATIO BAD STD FLANN LSH ENABLED & 0.24211 & 0.04491 & 1.04322\\ \hline
raw phishing COLORED PNG ORB 500 LEN MAX RATIO BAD STD FLANN LSH DISABLED & 0.24211 & 0.04502 & 1.05124\\ \hline
raw phishing COLORED PNG ORB 500 LEN MIN RATIO BAD STD FLANN LSH DISABLED & 0.24211 & 0.04556 & 1.05038\\ \hline
raw phishing COLORED PNG ORB 500 LEN MIN RATIO BAD STD BRUTE FORCE DISABLED & 0.23684 & 0.04937 & 1.31014\\ \hline
raw phishing COLORED PNG ORB 500 LEN MAX RATIO BAD STD BRUTE FORCE DISABLED & 0.23684 & 0.05403 & 1.20556\\ \hline
raw phishing COLORED PNG ORB 500 LEN MAX NO FILTER STD FLANN LSH ENABLED & 0.05789 & 0.04412 & 0.99475\\ \hline
raw phishing COLORED PNG ORB 500 LEN MAX NO FILTER STD FLANN LSH DISABLED & 0.03684 & 0.04388 & 0.99439\\ \hline
raw phishing COLORED PNG ORB 500 LEN MAX NO FILTER STD BRUTE FORCE DISABLED & 0.02105 & 0.05104 & 0.8506\\ \hline
raw phishing COLORED PNG ORB 500 LEN MIN NO FILTER STD FLANN LSH DISABLED & 0.0 & 0.0444 & 0.99507\\ \hline
raw phishing COLORED PNG ORB 500 LEN MIN NO FILTER STD BRUTE FORCE DISABLED & 0.0 & 0.08486 & 1.23023\\ \hline
raw phishing COLORED PNG ORB 500 LEN MIN NO FILTER STD FLANN LSH ENABLED & 0.0 & 0.04395 & 0.99463\\ \hline
\end{tabular}
\end{adjustbox}

\clearpage
\pagebreak
\section{Intersection matrix}
\label{app:A}

\begin{equation}
I_{ratio} = \frac{\sharp(Re\cap Gt)}{\sharp(Re)}
\label{formula:intersection}
\end{equation}
With $I_{ratio}$ being the intersection ratio, $Re$ being one algorithm output graph edges, $Gt$ being the ground truth graph edges or another algorithm output graph edges. The \textit{true-positive score} of an algorithm output is computed as the intersection ratio between one algorithm output and the ground truth graph.\\

Ground truth graph has \textit{clique} \footnote{Clique : subset of nodes of a graph where every two distinct nodes are adjacent}  \cite{hararyProcedureCliqueDetection1957} of similar nodes. It also has outliers which don't have any link with other nodes. This implies that algorithms which always give a matching picture to a given request picture (in other words algorithms which can't say "\textit{There is no match}") can't reach the maximum score. Some nodes will have edges that will never be "good" as their equivalent node in the \textit{ground truth} graph have no outbound edge. Some guess will always be wrong if a guess is forced.
Therefore, a normalization is applied (Table \ref{table:png_scores}) to raw true-positive rates as $normalized\_score = raw\_score/max\_score$.

\begin{figure}[!h]
\centering
\includegraphics[width=\textwidth]{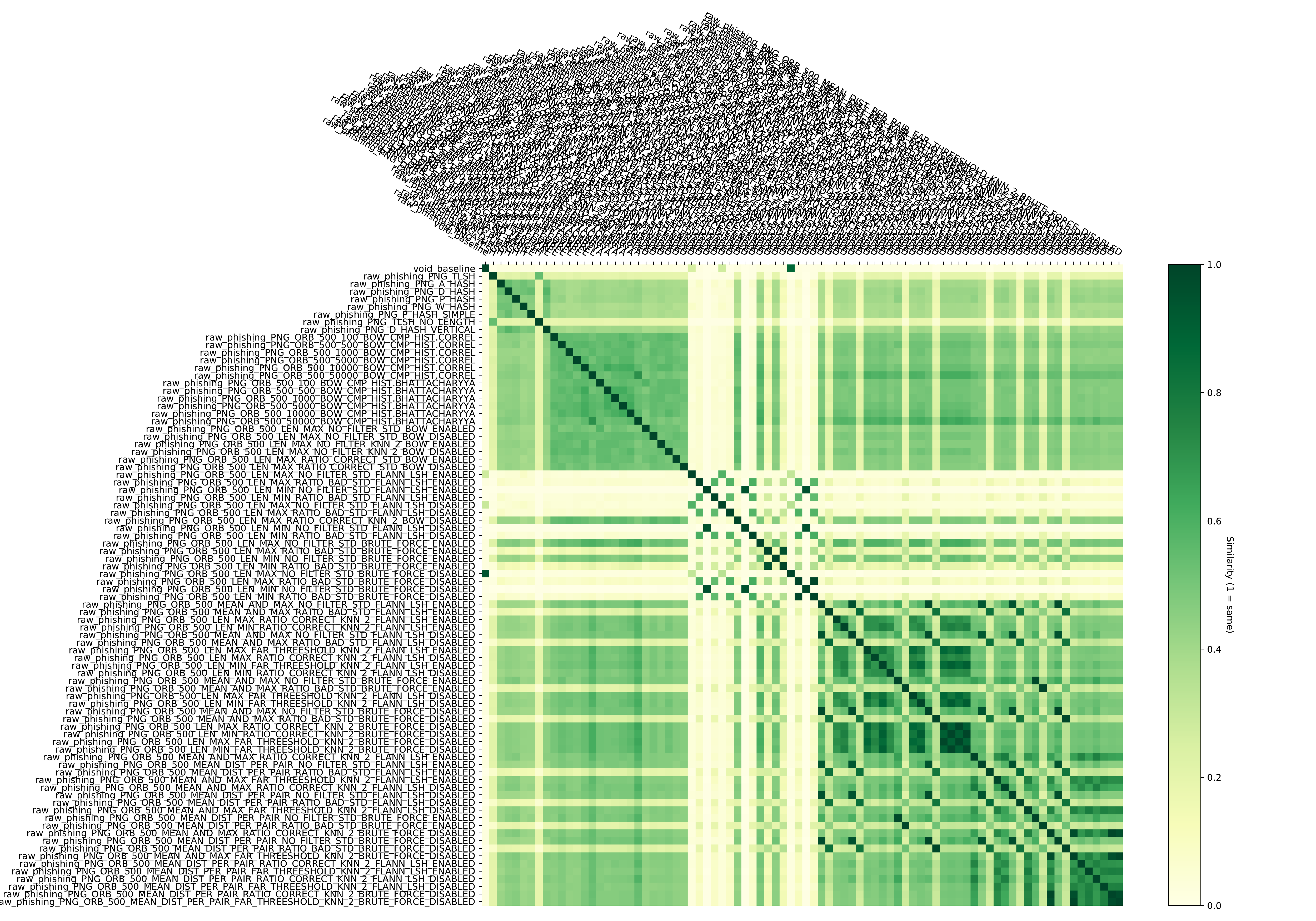}
\caption{Intersection matrix - Guess of rank 1 for each image comparison {\color{red}\textbf{CONCATENE AND MAKE IT MORE READABLE : FOR NEXT RELEASE}} }
\label{fig:intersection0}
\end{figure}

For all given algorithm, we compute the intersection of their outputs in pairs, thanks to Formula \ref{formula:intersection}. The matrix is presented in Figure \ref{fig:intersection0}.
The larger the intersection between an Algorithm A (y-axis) and an Algorithm B (x-axis), the "\textit{more similar}" their output will be, the darker the square of their intersection will be. 

We see that numerous ORB-configuration are similar (homogeneous dark bottom-right square), while most hash-based algorithms have similar outputs (somewhat homogeneous mid-dark top-left square). We also see that hash-based algorithms and ORB-based configuration have quite different outputs (homogeneous light areas). Therefore, combination of hash-based algorithms and ORB-based configuration may be complementary.

The evaluation had been conducted on the rank-1 matching guess provided for each algorithm for each request picture.

\end{document}